\documentclass[showpacs,superscriptaddress,floatfix,amsmath,amssymb,twocolumn,pra]{revtex4-1}
\usepackage{bm}
\usepackage{graphicx}
\usepackage{color}
\usepackage[normalem]{ulem}
\usepackage{epstopdf}
\usepackage{soul}
\usepackage{url}
\usepackage[normalem]{ulem}

\usepackage{hyperref}
\hypersetup{pdftex,colorlinks=true,allcolors=blue,bookmarksnumbered=true}
\usepackage{hypcap}

\def\sgn{{\rm sgn}}
\def\Tr{{\rm Tr}}
\def\Re{{\rm Re}}

\def\ep{\varepsilon}

\begin{document}

\title {Preparing quasienergy states on demand: A parametric oscillator}
\author{Yaxing Zhang}
\affiliation{Department of Physics, Yale University, New Haven, CT 06511, USA}
\author{M. I. Dykman}
\affiliation{Department of Physics and Astronomy, Michigan State University, East Lansing, MI 48824, USA}


\begin{abstract}
We study a nonlinear oscillator, which is parametrically driven at a frequency close to twice its eigenfrequency. By judiciously choosing the frequency detuning and linearly increasing the driving amplitude, one can prepare any even quasienergy  state starting from the oscillator ground state. Such state preparation is effectively adiabatic. We find the Wigner distribution of the prepared states. For a different choice of the frequency detuning, the adiabaticity breaks down, which allows one to prepare on demand a superposition of quasienergy states using Landau-Zener-type transitions. We find the characteristic spectrum of the transient radiation emitted by the oscillator after it has been prepared in a given quasienergy state.

\end{abstract}

\maketitle

\section{Introduction}

Periodically driven quantum systems are described by quasienergy (Floquet) states, which are a time-domain analog of Bloch states in spatially periodic systems \cite{Shirley1965,Zeldovich1967,Ritus1967,Sambe1973}. The new physics associated with quasienergy states has been attracting much interest recently. Examples include topological Floquet states, artificial gauge fields, and new many-body phases \cite{Kitagawa2010,Lindner2011,Goldman2014,Bukov2015,Peano2016,Khemani2016,Keyserlingk2016,Khemani2016a,Zhang2016,Choi2016,Bairey2017}.

Preparation of Floquet states is often discussed in the adiabatic framework assuming that the periodic field is slowly turned on, cf.~\cite{DAlessio2015,Heinisch2016,Weinberg2016,Ho2016} and references therein. The analysis for many-body systems is complicated by the effect of heating, and much progress has been made by studying 
systems that display many-body localization, as it may alleviate the heating. Recently, adiabatic state preparation was considered also for a parametrically driven nonlinear oscillator \cite{Goto2016,Puri2016}. In contrast to many-body systems, the energy spectrum here is discrete, which simplifies the problem. However, a potential complication, and also potentially new and interesting features stem from the fact that the quasienergy levels for weak driving  can display degeneracy, or a specific type of degeneracy, which we call the reduced-band (RB) degeneracy.

The goal of this paper is to study preparation of quasienergy states in a small quantum system in the case where the quasienergy states can display degeneracy or the RB degeneracy for weak driving. In optics terms, this case corresponds to either a multiphoton resonance or a subharmonic resonance, where the distance between the energy levels of the system is close to either a multiple or a fraction of the radiation frequency multiplied by $\hbar$.  Multiphoton resonance leads to Rabi oscillations described in Ref.~\onlinecite{Larsen1976} for a nonlinear oscillator using perturbation theory. In terms of the Floquet states, when the driving frequency is close to the oscillator eigenfrequency, such oscillator can display simultaneous multiple anticrossing of the quasienergy levels \cite{Dykman2005}.

We will use as a model a driven quantum oscillator. Such model is interesting as it describes a broad range of physical systems, from molecular vibrations \cite{Larsen1976} to the modes of nonlinear optical and microwave cavities to Josephson junctions \cite{Dykman2012}. Here we study the features of the Floquet dynamics that emerge when an oscillator is driven parametrically and the drive frequency $\omega_F$ is close to twice the oscillator eigenfrequency.

To explain how the multiphoton and subharmonic resonances are seen in the quasienergy spectrum, we note that  quasienergies of a system $\ep_n$ and the quasienergy level spacing $\ep_n-\ep_m$ are defined modulo $\hbar \omega_F$. In the limit of zero driving $\ep_n-\ep_m$ is simply related to the spacing ${\cal E}_n-{\cal E}_m$ of the corresponding energy levels of the system, $\ep_n -\ep_m=({\cal E}_n-{\cal E}_m)\!\!\!\mod(\hbar\omega_F)$.  The standard multiphoton resonance for weak driving occurs if ${\cal E}_n-{\cal E}_m$ is a multiple of $\hbar\omega_F$, and then $\ep_n-\ep_m=0$ for a given pair of states $(n,m)$, i.e., the quasienergies are degenerate. In contrast, in the case of a subharmonic resonance, ${\cal E}_n-{\cal E}_m$ can be a fraction of $\hbar\omega_F$. In particular, for the parametric resonance in an oscillator one can have $|{\cal E}_n-{\cal E}_{n+1}|=\hbar\omega_F/2$ (a more general resonant condition is discussed below, cf. Fig.~\ref{fig:zerodrive}). In this case $|\ep_n-\ep_{n+1}|=\hbar\omega_F/2$. This is the RB degeneracy, as the quasienergies would coincide if they were defined modulo $\hbar\omega_F/2$. Such degeneracy is nontrivial, since if the system is prepared in a superposition of the RB-degenerate states, it displays period doubling: the state is reproduced (up to a trivial phase factor) after twice the driving period, rather than after one period.

In what follows we show that, by slowly turning on resonant parametric drive, it is possible to prepare on demand various quasienergy states starting from the ground state of the oscillator ($n=0$).  Importantly, this can be done in a finite time and with high accuracy without using special pulse-shaping techniques, but just by increasing the amplitude of the drive linearly in time. Such scenario is easy to implement in the experiment. We also study preparation of a superposition of quasienergy states starting from the ground state. Such preparation can be accomplished using non-adiabatic transitions for the driving frequency $\omega_F$ tuned close to multiphoton resonance, so that $\ep_m-\ep_0$ is small for the targeted $m$. Again, this relies on a simple linear increase of the driving amplitude. However, the nonadiabatic dynamics in this case turns out to be different from the conventional Landau-Zener dynamics. 

The paper is organized as follows. In Sec.~II, we present the model of a parametric nonlinear oscillator and discuss its quasienergy spectrum. We show the evolution of the spectrum with the varying driving frequency in the limit of zero drive amplitude and the occurrence of the degeneracy and the RB degeneracy of the quasienergy levels as the system goes through multiphoton or subharmonic resonance. In Sec.~III, we present the Wigner distribution for the quasienergy states prepared from the oscillator ground state by slowly ramping up the amplitude of the driving in the absence of degeneracy. We demonstrate the possibility to prepare a Floquet state ``on demand" and the rich structure of its  Wigner distribution. The only constraint is that the resulting Floquet states are ``even" with respect to inversion in phase space. In Sec.~IV, we consider preparation of a superposition of two quasienergy states via a non-adiabatic transition when the system is close to degeneracy for weak field. In Sec.~V, we briefly discuss the adiabaticity in the presence of dissipation. In Sec.~VI we study fluorescence of the oscillator driven into a Floquet state, and in particular the characteristic transient spectrum of the fluorescence. Sec.~VI contains concluding remarks.

\section{RWA Hamiltonian and quasienergy spectrum}
\label{sec:RWA}

The Hamiltonian of a weakly nonlinear parametric oscillator with coordinate $q$ and momentum $p$ has the form
\begin{equation}
H(t)= \frac{p^2}{2}+\frac{1}{2}q^2[\omega_0^2+F\cos(\omega_Ft)] + \frac{\gamma}{4}q^4.
\label{eq:Hamiltonian}
\end{equation}
We assume that the driving amplitude $F$ and the nonlinearity are comparatively small, $F, \gamma \langle q^2 \rangle \ll \omega_0^2$, and the driving frequency $\omega_F$ is close to resonance, $|\omega_F-2\omega_0| \ll \omega_0$; without loss of generality, we consider $F, \gamma>0$. A quantum parametric oscillator described by Eq.~(\ref{eq:Hamiltonian}) has been realized in various platforms, from optical and microwave cavities to nanomechanical systems, cf. \cite{Nabors1990,Wilson2010,Dykman2012,Lin2014}. 

For a periodically modulated quantum system, there exists a complete set of solutions  to the Schr\"odinger equation called Floquet states, which are eigenfunctions of the operator $T_{t_F}$ of time translation by the modulation period ${}t_F$,
\begin{equation}
\psi_\ep(t) = e^{-i\ep t/\hbar} u_\ep(t), u_\ep(t+{}t_F) =u_\ep(t),
\label{eq:Floquetdef}
\end{equation}
Parameter $\ep$ is called quasienergy or Floquet eigenvalue. For the parametric oscillator with Hamiltonian (\ref{eq:Hamiltonian}), ${}t_F = 2\pi/\omega_F$.

A standard procedure to find quasienergy states and quasienergies is to plug the solution Eq.~(\ref{eq:Floquetdef}) into the Schr\"odinger equation, and then solve the resulting equation for $u_\ep(t)$ using Fourier series expansion; see Appendix. For a driven oscillator, a much simpler way to find quasienergies is to go to the rotating frame at frequency $\omega_F/2$ by applying the standard unitary transformation $U(t)=\exp[-i\omega_F a^\dagger at/2]$, where $a$ and $a^\dagger $ are the oscillator ladder operators. In the rotating wave approximation (RWA) we disregard fast oscillating terms in the transformed Hamiltonian $U^\dagger HU - i\hbar U^\dagger \dot U$, which gives the RWA Hamiltonian 
\begin{equation}
\label{eq:H_eff}
H_{\rm RWA} = -\hbar\delta\omega_F  \hat n + \frac{\hbar V}{2}(\hat n^2 + \hat n) + \frac{\hbar \tilde F}{2}(a^2+a^{+2})
\end{equation}
where $\hat n = a^\dagger a$, $\delta\omega_ F = \omega_F/2-\omega_0$ is the detuning frequency, $\tilde F = F/4\omega_0$, and $V = 3\gamma \hbar/4\omega_0^2$. 

The Hamiltonians $H$ and $H_{\rm RWA}$ commute with the parity operator $\hat P =  \exp(-ia^\dagger a \pi)$ \cite{Haroche2006} that transforms $q\to -q,p\to -p$. Therefore, an eigenstate  $\phi_E$ of $H_{\rm RWA}$ has definite parity $P_E = \pm1$; here $E$ is an eigenvalue of $H_{\rm RWA}$, which can be called the RWA energy, $H_{\rm RWA}\phi_E = E\phi_E$.  As a consequence, the corresponding time dependent state in the lab frame $\Phi_E(t)\equiv \exp(-iEt/\hbar) U(t) \phi_E$ is a Floquet state of Eq.~(\ref{eq:Floquetdef}). The quasienergy $\ep$  and the periodic factor in the Floquet wave functions $u_\ep$ are immediately expressed in terms of the RWA energy $E$ and the eigenfunction $\phi_E$,
\begin{align}
&\ep= [E +(1-P_E )\hbar \omega_F/4]{\rm mod} (\hbar\omega_F),\nonumber\\
&u_\ep(t) = 
\exp[i(1-P_E)\omega_F t/4] U(t)\phi_E, \nonumber\\
&\text{where } P_E = \left\{
\begin{array}{ll}
-1, &\phi_E \text{ is odd}\\
1, &\phi_E \text{ is even}.
\end{array}\right. 
\label{eq:RWAFloquetrelation}
\end{align}
The aforementioned RB degeneracy where the quasienergies differ by $\hbar\omega_F/2$ occurs if $H_{\rm RWA}$ has degenerate states $\phi_E$. Such degeneracy is possible for a parametric oscillator for a finite driving amplitude \cite{Marthaler2007a}. A driven oscillator also provides a platform for investigating more complicated cases of RB degeneracy \cite{Zhang2017}.

\begin{figure}[ht]
\includegraphics[width = 7truecm]{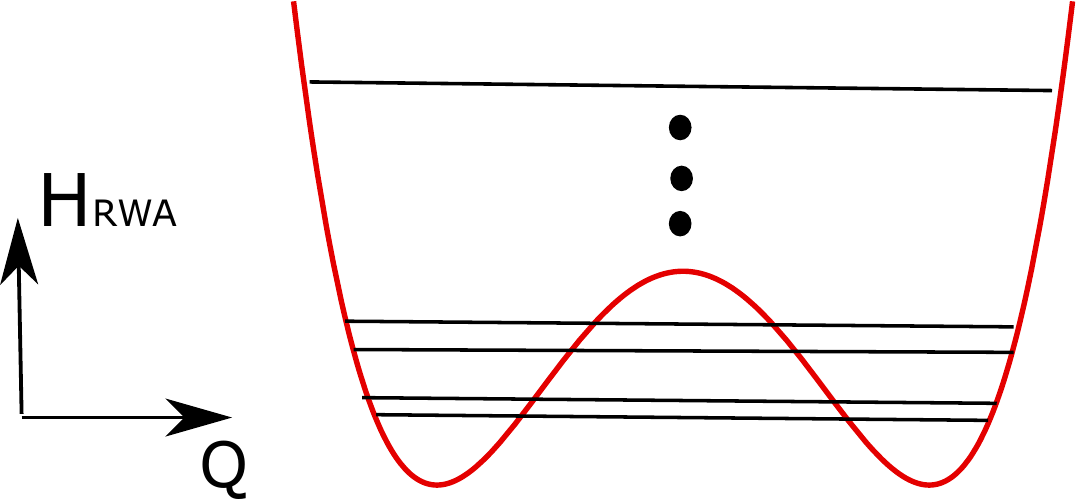}
\caption{The cross-section of the RWA Hamiltonian function $H_{\rm RWA}(Q,P)$ given by Eq.(\ref{eq:g_function}) by the plane $P=0$ and the RWA energy levels.}
\label{fig:schematic}
\end{figure} 

The understanding of the spectrum of $H_{\rm RWA}$ can be gained by looking at the Hamiltonian function $H_{\rm RWA}$ in the phase space of the oscillator in the rotating frame, i.e., by writing $H_{\rm RWA}$ in terms of the scaled quadratures $P$ and $Q$ defined as $Q =  i(a-a^\dagger )\sqrt{\lambda/2}, P = (a^\dagger +a)\sqrt{\lambda/2}$. Here, $\lambda = V/2\tilde F$ is the dimensionless Planck constant. In these variables 
\begin{align}
\label{eq:g_function}
&H_{\rm RWA}(Q,P) = (F^2/6\gamma) g(Q,P), \nonumber\\ 
&g(Q,P) =\frac{1}{4}(P^2+Q^2)^2  -\frac{1}{2}\mu(P^2+Q^2) + \frac{1}{2}(P^2-Q^2), 
\end{align}
where  $\mu =2 \omega_F(\delta\omega_F)/F$ \cite{Marthaler2007a}. The eigenstates of the Hamiltonian $H_{\rm RWA}$ can be written in the $Q$-basis, $\phi_E\equiv \phi_E(Q)$. The parity operator $\hat P$ is then the inversion operator, $\hat P\phi_E(Q) = \phi_E(-Q)$. 

For $\mu+1>0$, function $H_{\rm RWA}(Q,P)$ has two minima located at $P=0, Q = \pm\sqrt{\mu+1}$. 
Function $H_{\rm RWA}(Q,P=0)$ is shown in Fig.~\ref{fig:schematic}. For sufficiently strong driving, where the two wells become deep and well-separated, the low-lying eigenstates of $H_{\rm RWA}$ are symmetric or anti-symmetric superpositions of intra-well states.    

\begin{figure}[ht]
\includegraphics[width = 8truecm]{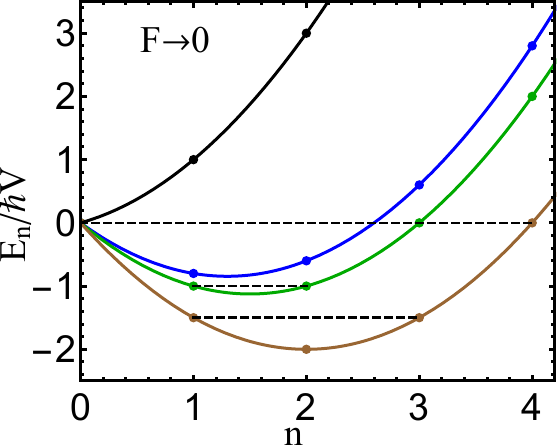}
\caption{RWA energies $E_n$ in the limit $F\rightarrow 0$. The curves from top down correspond to $\delta\omega_F/V =$ 0 (black),  1.8 (blue), 2 (green), and 2.5 (brown). The solid lines are guides for eyes;  the values of the energies are indicated by the dots, which refer to integer values of $n$.  The dashed lines are intended to show the degeneracy: $E_0 = E_3, E_1=E_2$ (green curve); $E_0 = E_4$, $E_1=E_3$ (brown curve). }
\label{fig:zerodrive}
\end{figure}

In the opposite limit of weak driving, $F \rightarrow 0$, the Hamiltonian $H_{\rm RWA}$, Eq.~(\ref{eq:H_eff}), is trivially diagonalized in the basis of the oscillator Fock states. What is interesting, however, is that the order of the RWA eigenstates in the rotating frame can be  changed compared to the order of the Fock states in the laboratory frame. From Eq.~(\ref{eq:H_eff}), for $F=0$ the eigenvalues $E_n$ of $H_{\rm RWA}$ can be written in a suggestive form
\begin{equation}
\label{eq:zerodrive}
E_n = \bar E_n - \bar E_0,\quad 
\bar E_n=\frac{1}{2}\hbar V\left(n+\frac{1}{2}-\frac{\delta\omega_F}{V}\right)^2.
\end{equation}
From Eq.~(\ref{eq:zerodrive}), $E_n$ considered as a continuous function of $n$ is a simple parabola with a minimum at $n=\delta\omega_F/V - 1/2$; see Fig.~\ref{fig:zerodrive}. For $\delta\omega_F/V < 1/2$, $E_n$ quadratically increases with the increasing $n$; see the top line in Fig.~\ref{fig:zerodrive}. However, as the ratio $\delta\omega_F/V$ increases, $E_n$ bends over and has a minimum at some positive $n$. Of course, the actual RWA energies are determined by $E_n$ with integer $n$. When $\delta\omega_F/V>1$, the state with the lowest $E_n$ is no longer the Fock state $|0\rangle$. For instance, for $\delta\omega_F/V = 1.8$ (blue dots, which lie on the second from top line in Fig.~\ref{fig:zerodrive}), this state is $|1\rangle$.

The reordering of the quasienergy states described by Eq.~(\ref{eq:zerodrive}) is essential for preparing quasienergy states on demand. Indeed, if the oscillator is initially in the ground state, then by tuning the driving frequency and increasing the driving strength, we make this state an arbitrary even in $Q$ quasienergy state, i.e., an arbitrary superposition of Fock states $|m\rangle$ with even $m$. We also note that, for certain values of $\delta\omega_F/V$, there can be degenerate RWA levels (the green and brown dots, which lie on the two lowest curves  in Fig.~\ref{fig:zerodrive} and are connected by dashed lines). We will discuss such degeneracy later in details. 

\begin{figure}[ht]
\includegraphics[width = 9truecm]{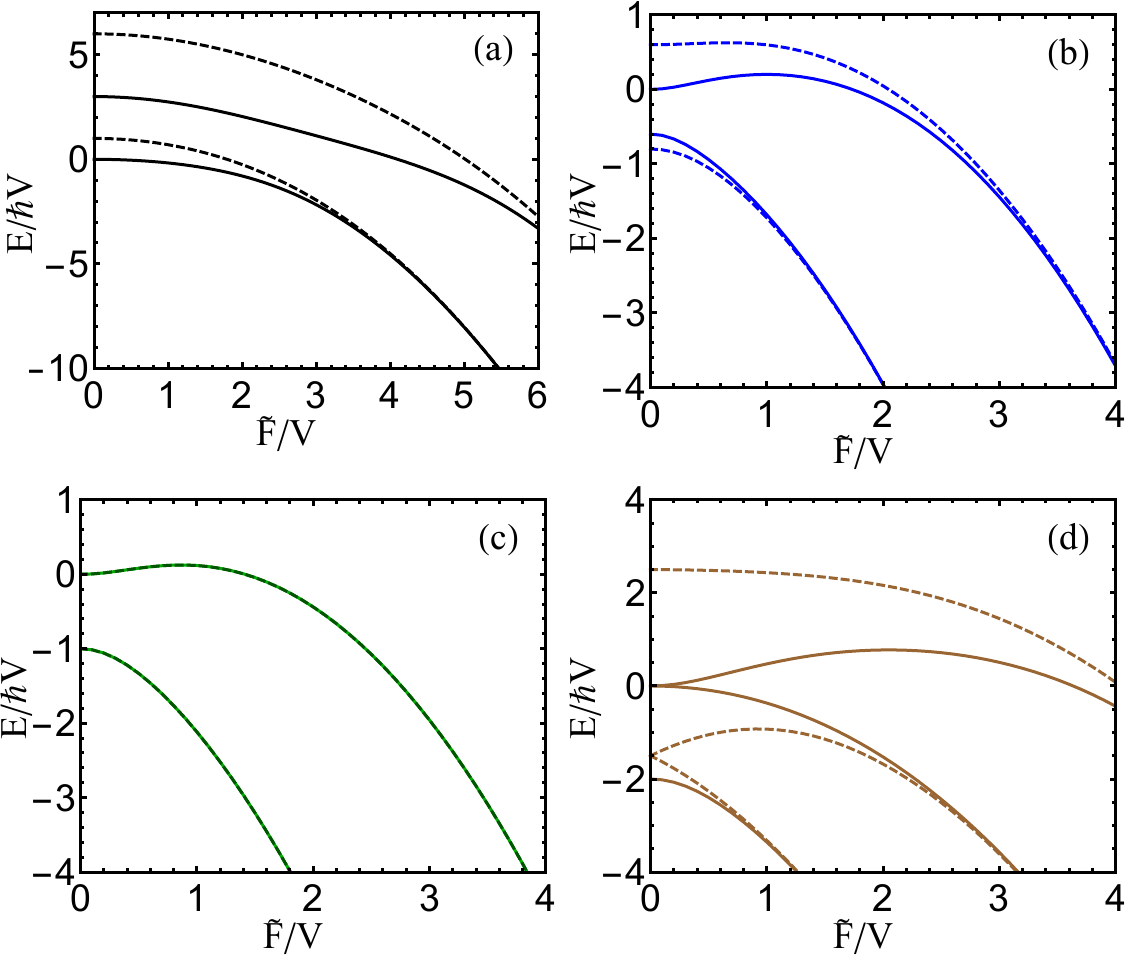}
\caption{Evolution of the RWA energy spectrum with the  increasing driving amplitude $F$  for $\delta\omega_F/V = 0 (a), 1.8(b), 2(c), 2.5(d).$ The solid and dashed lines refer to the RWA energy levels of even and odd parity, respectively. In panel (c), the solid and dashed line coincide.} 
\label{fig:quasienergy_spectrum}
\end{figure} 

The driving mixes Fock states with the same parity. The evolution of the RWA spectrum with the increasing $F$ is shown in Fig.~\ref{fig:quasienergy_spectrum} for different values of the detuning $\delta\omega_F/V$. A common trend is that RWA energy levels of the same parity repel each other, whereas neighboring levels of opposite parity attract each other and form pairs for large $\tilde F/V$. As mentioned above, such pairs for large $\tilde F/V$ are even and odd superposition of ``intra-well'' states of $H_{\rm RWA}$. The distance between the states within the pairs is determined by interwell tunneling \cite{Marthaler2007a}. 

\subsection{Special features of the RWA spectrum}

We find that, somewhat counterintuitively, the RWA levels do not cross each other as $F$ changes. Therefore any gaps that are present at $F\rightarrow 0$ will remain open for any finite $F$. For instance, Figs.~\ref{fig:quasienergy_spectrum}a and b refer to the cases where the Fock state $|0\rangle$ is the first and the third lowest RWA eigenstate at $F\rightarrow 0$, respectively. As $F$ increases, it remains the first and the third lowest RWA eigenstate. Such non-crossing feature will be important for the preparation of quasienergy states by slowly turning on the driving.

A remarkable feature of the RWA spectrum is that, when the ratio $\delta\omega_F/V$ is a positive integer, there is a set of simultaneously doubly-degenerate levels of opposite parity regardless of the value of $F$. For $F\rightarrow 0$, this can be readily seen from Eq.~(\ref{eq:zerodrive}) (cf. \cite{Dykman2005} where a similar feature was found in the case of the driving at frequency close to $\omega_0$). When $\delta\omega_F/V = k, k = 1,2,3..$, the minimum of $E_n$ as a continuous function of $n$ is reached at half odd integer $n =  k-1/2$. Since $E_n$ is a symmetric function of $n$ with respect to the minimum, the levels separated by $\Delta n = 2m+1$ are degenerate, that is, $E_{k+m}=E_{k-(m+1)}$, for $m = 0,1..,k-1$. The green curve in Fig.~\ref{fig:zerodrive} (the third from the top) refers to the case $k=2$, where the degeneracy condition is met. 

The degeneracy of the RWA energy levels persists for nonzero $F$, as can been seen in Fig.~\ref{fig:quasienergy_spectrum}c. At weak driving, this follows from the perturbation theory. To the second order in $F$, the correction $\delta E_n$ to  $E_n$ is
\begin{equation}
\delta E_n =-\hbar V \frac{\tilde F^2}{4V^2}\frac{2\bar E_n/\hbar V -(\delta\omega_F/V)^2-3/4}{2\bar E_n/\hbar V-1}
\end{equation} 
The dependence of $\delta E_n$ on the level number $n$ is exactly the same as that of $E_n$. cf. Eq.~(\ref{eq:zerodrive}). Therefore, if $E_n = E_{n'}$, then $\delta E_n = \delta E_{n'}$. Note that the perturbation theory still applies even if there are degenerate levels of opposite parity since there is no coupling between them.  At strong driving, such degeneracy corresponds to the vanishing of tunnel splitting found in Ref.~\cite{Marthaler2007a}.

For the special case $\delta\omega_F/V = 1$, $H_{\rm RWA}$ can be factored \cite{Puri2016}, 
\begin{equation}
H_{\rm RWA} = \frac{\hbar V}{2}\left(a^{+2} + \frac{\tilde F}{V}\right)\left(a^2 + \frac{\tilde F}{V}\right) - \frac{\hbar \tilde F^2}{2V}. \nonumber
\end{equation}
In this case the coherent states $| \pm \alpha\rangle$, $\alpha = \sqrt{-\tilde F/V}$, are exact degenerate eigenstates of $H_{\rm RWA}$ for arbitrary driving strength. However, no such eigenstates are known for other values of $\delta\omega_F/V$.

If the ratio $\delta\omega_F/V$ is a half-integer, $\delta\omega_F/V=(2k+1)/2, k = 1,2,3,...$, the minimum of function $E_n$  for $F\rightarrow 0$ is reached at integer $n = k$. Again, due to the parabolic dependence of $E_n$ on $n$, levels $E_{k \pm m}$ are degenerate for $m=1,2,...,k$. For instance, the lowest (brown) curve in Fig.~\ref{fig:zerodrive} refers to the case $k = 2$. The degeneracy of the levels of the same parity occurs when the driving frequency equals to one of the transition frequencies of the undriven oscillator. This can be seen by rewriting $E_n$ as $E_n = -n\hbar\omega_F/2 + \mathcal E_n$, where $\mathcal E_n = n\hbar \omega_0 + \hbar V n(n+1)/2$ is the $n$th energy level of the oscillator in the absence of driving. Clearly, the degeneracy condition $E_{k+m}=E_{k-m}$ is equivalent to $\mathcal E_{k+m}- \mathcal E_{k-m} = m\omega_F$, which is the m-photon resonance condition for transition from $\mathcal E_{k-m}$ to $\mathcal E_{k+m}$. The degeneracy is lifted at finite $F$ due to level repulsion, as shown in Fig.~\ref{fig:quasienergy_spectrum}d.

\section{Effectively-adiabatic 
preparation of quasienergy states and the Wigner distribution}
\label{sec:Wigner}

\begin{figure}[ht]
\includegraphics[width = 9truecm]{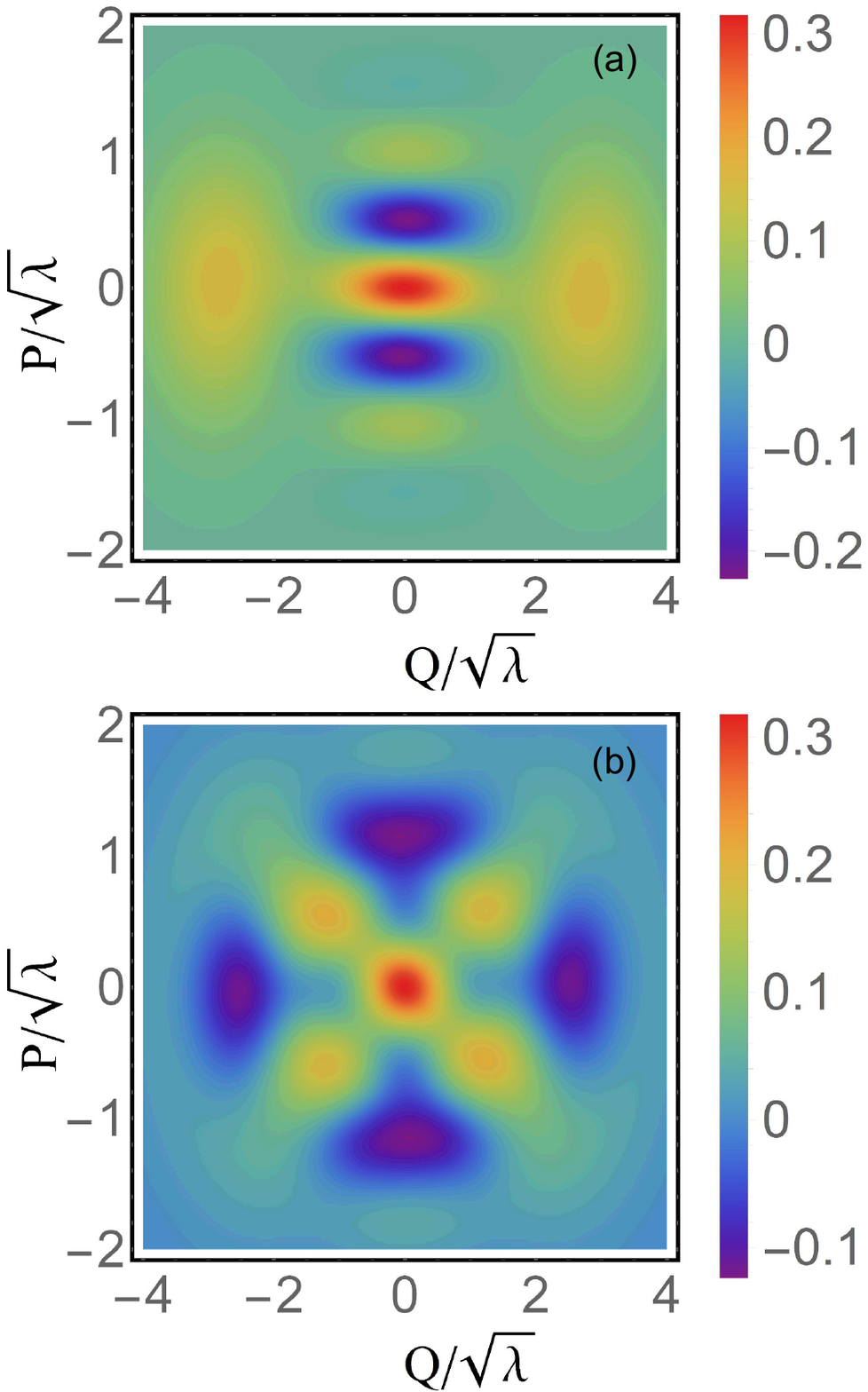}
\caption{The density matrix of the oscillator at time $\tilde F_{\rm final}/s_0$ in the Wigner representation for a linear ramp, $\tilde F(t) = s_0t$. The oscillator is in state $|\phi(Q)\rangle$, which  is obtained from the time-dependent Schr\"odinger equation (\ref{eq:Schrodinger}) assuming that $\phi(Q) = |0\rangle$ for $t=0$. The parameters are: (a) $\delta\omega_F/V = 0, \tilde F_{\rm{final}}/V = 5, s_0/V^2 = 1$, and  (b) $\delta\omega_F/V = 1.8, \tilde F_{\rm{final}}/V = 3, s_0/V^2 = 0.06$. }
\label{fig:Wigner_function}
\end{figure} 

The observation that the quasienergy levels of the same parity do not approach each other with the increasing field $F$ is critical for state preparation. It allows one to prepare a quasienergy state by slowly turning on the field, provided the states are non-degenerate for $F\to 0$.

We consider ramping up the driving amplitude $\tilde F$ linearly with speed $s$  starting at $t=0$, $\tilde F(t) = s_0t$. If $\sqrt{s_0}$ is small compared to $\omega_0$, the time evolution of the oscillator wave function $\phi(t)$ can be described in the RWA,
\begin{equation}
\label{eq:Schrodinger}
i\hbar \partial_t \phi(t) = H_{\rm RWA}(t)\phi(t).
\end{equation}
We will solve this equation assuming that initially, for zero driving, the system is in the ground state of the oscillator, $\phi(Q,t=0)=|0\rangle$. 

The results of the numerical solution of Eq.~(\ref{eq:Schrodinger}) are illustrated in Fig.~\ref{fig:Wigner_function}. The values of $\delta\omega_F/V$ were chosen in such a way that, in one case ($\delta\omega_F=0$), the state of the system remains close to the eigenstate of $H_{\rm RWA}$ with the lowest eigenvalue $E_n$, whereas in the other case ($\delta\omega_F/V=1.8$) it is close to the third lowest-$E_n$ state, cf. Fig.~\ref{fig:quasienergy_spectrum}(b). The quality of the adiabatic approximation for the chosen parameters can be characterized by the inner product of the state $\phi(Q)$ at the end of ramp-up and the corresponding stationary RWA eigenstate $\phi_E(Q)$ calculated for $\tilde F=\tilde F_{\rm final}$. This inner product is 0.997 and 0.98 for the cases shown in Fig.~\ref{fig:Wigner_function}a and Fig.~\ref{fig:Wigner_function}b, respectively, which shows that the adiabatic approximation is very good.

The final value of the field amplitude $\tilde F_{\rm final}$ in Fig.~\ref{fig:Wigner_function} refers to the case where the Hamiltonian function $H_{\rm RWA}(Q,P)$, Eq.~(\ref{eq:g_function}), has a pronounced double-well structure, cf. Fig.~\ref{fig:schematic}. For $\delta\omega_F=0$, the state $\phi(Q)$ is well described by a symmetric superposition of the lowest  intra-well states in Fig.~\ref{fig:schematic}, $\phi(Q)= (\phi_L + \phi_R)/\sqrt{2}$ where $\phi_L$ and $\phi_R$ refer to the left and right well, respectively. Near their maxima, functions $\phi_{L,R}$ are given by squeezed coherent states with equal amplitude and opposite phases, $\phi_{L,R} \propto \exp[-(Q\pm Q_0)^2/2\lambda\eta]$ where  $Q_0 = \sqrt{\mu+1}$ is the position of the right well and $\eta = 1/\sqrt{\mu+1}$ characterizes the state squeezing, see Appendix~\ref{App:semiclassical}. The adiabatic preparation of such ``cat" state has been discussed in Refs.~(\cite{Goto2016, Puri2016}). 

In contrast, for the case in Fig.~\ref{fig:Wigner_function}b, the driving brings the system to an excited state of $H_{\rm RWA}$. The state $\phi(Q)$ for $t=\tilde F_{\rm{final}}/s_0$ is no longer a superposition of the lowest intra-well states but, for the chosen $\delta\omega_F/V$, the superposition of the second lowest intra-well states, $\phi(Q) =  (\phi_L' + \phi_R')/\sqrt{2}$. Near their maxima, functions $\phi_{L,R}'$ are well described by a displaced and squeezed Fock state $|1\rangle$: $\phi_{L,R}' \propto (Q\pm Q_0)\exp[-(Q\pm Q_0)^2/2\lambda\eta].$ Since the RWA energy levels for small $F$ in this case are closer than for $\delta\omega_F=0$, in particular the Fock states $|0\rangle$ and $|2\rangle$ have close RWA energies,  we had to use a much slower increase of the driving amplitude to attain high fidelity of the prepared large-$F$ state.

\section{Preparing a superposition of quasienergy states nonadiabatically}
\label{sec:non-adiabatic}
As the driving amplitude $F$ is ramped up, the non-adiabaticity can mix quasienergy states of the same parity. The mixing is particularly strong if the quasienergy gap that separates the states is small. As shown in Sec.~\ref{sec:RWA}, this gap is controlled by the driving frequency. In this section, we consider a situation where two nearest quasienergy states of the same parity have close quasienergies for $F\to 0$, whereas the quasienergies of other states are significantly different, so that mixing with these other states can be disregarded for slowly varying $F(t)$. We show that, by ramping up the driving amplitude linearly in time, we can prepare, with high accuracy, a desired coherent superposition of the chosen two quasienergy states. 

We assume that the states with close quasienergies for $F\to 0$ are $|n-1\rangle $ and $|n+1\rangle$, which means that $\delta\omega_F/V\approx n+1/2$. As the drive is ramped up, these states are mixed with each other. Concurrently, they are mixed with other states of the same parity. However, this mixing is nonresonant and therefore is weaker. 

\begin{figure}[ht]
\includegraphics[width=7truecm]{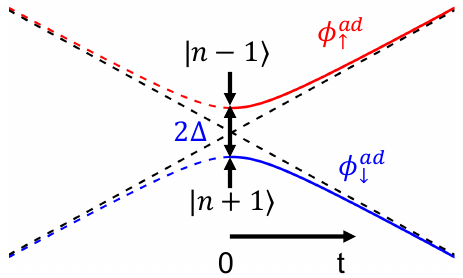}
\caption{A schematic of two resonating quasienergy states in the adiabatic picture. The plot refers to $\Delta >0$.}
\label{fig:semi-LandauZener}
\end{figure}

The picture of the state evolution is as follows. The resonant mixing leads to a redistribution of the initial population between the resonating states and to a separation of their quasienergies already for a comparatively weak field, see Fig~\ref{fig:semi-LandauZener}. The increase of the field afterwards does not change the state populations, even though it modifies the states by increasingly strongly admixing them to other states of the same parity.

To describe the initial stage of the evolution we project the Hamiltonian $H_{\rm RWA}$ onto the subspace formed by the states $|n-1\rangle$ and $|n+1\rangle$, subtract the mean RWA energy $(E_{n+1} +E_{n-1})/2$, and disregard the coupling to other states. Then the Hamiltonian becomes
\begin{equation}
\label{eq:twobytwo}
H_{\rm RWA}(t) = \hbar \left (
\begin{matrix}
\Delta & \nu(t)\\
\nu(t) & -\Delta 
\end{matrix} \right ),
\end{equation}
where $\Delta = (E_{n-1}-E_{n+1})/2\hbar$, $\nu(t) =  \sqrt{n(n+1)}\tilde F(t)$. For a field that linearly increases in time $\nu(t) = s{}t$. 

It is convenient to re-write the Hamiltonian (\ref{eq:twobytwo}) in the conventional form used in the analysis of the Landau-Zener tunneling. Making a unitary transformation $U_\sigma = (1/\sqrt{2})(\sigma_z + \sigma_x)$ ($\sigma_{x,z}$ are Pauli matrices), we obtain 
\begin{equation}
\label{eq:LZ}
U_\sigma^\dagger H_{\rm RWA}U_\sigma = H_{\rm LZ} = \hbar \left (
\begin{matrix}
\nu(t) & \Delta\\
\Delta & -\nu(t) 
\end{matrix} \right ).
\end{equation}
Note that the vectors $\left(\begin{matrix}1\\0\end{matrix}\right)$ and $\left(\begin{matrix}0\\1\end{matrix}\right)$ for the Hamiltonian (\ref{eq:LZ}) are, respectively, the wave functions  $(|n-1\rangle + |n+1\rangle)/\sqrt{2}$ and $(|n-1\rangle - |n+1\rangle)\sqrt{2}$.

The only difference of the evolution of the states we consider here from the standard Landau-Zener scenario is that the initial condition for the Schr\"odinger equation $i\hbar\dot \phi (t)= H_{\rm LZ}\phi(t)$ is set for $t=0$ and the problem is considered on the semi-axis $t\geq 0$. It is convenient to seek the wave function as $\phi(t) = (1/\sqrt{2})\sum_{\alpha=\pm}C_\alpha(t)[|n-1\rangle + \alpha|n+1\rangle]$. We will be interested in the solution that corresponds to the initial condition where the smaller-$n$ state is occupied while the larger-$n$ state is empty, $C_+(0) =C_-(0)=1/\sqrt{2}$. As in the Landau-Zener problem, the solution to the Schr\"odinger equation can be expressed in terms of the parabolic cylinder functions; see Appendix~\ref{app:Landau_Zener}.

In Fig.~\ref{fig:LandauZener}, we show the result for the coefficient $C_\uparrow(t)$, which is equal to the projection $\langle \phi^{\rm ad}_\uparrow(t)|\phi(t)\rangle$ of the wave function $\phi(t)$ on the upper branch (the higher energy branch in Fig.~\ref{fig:semi-LandauZener}) of the adiabatic solutions $\phi_{\uparrow,\downarrow}(t)$ of the Schr\"odinger equation, $H_{\rm LZ}\phi^{\rm ad}_{\uparrow,\downarrow}(t) =\pm  [ \nu^2(t)+\Delta^2 ]^{1/2}\phi^{\rm ad}_{\uparrow,\downarrow}(t)$. The result is in full agreement with the numerical solution of the Schr\"odinger equation.
\begin{figure}[ht]
\includegraphics[width=8truecm]{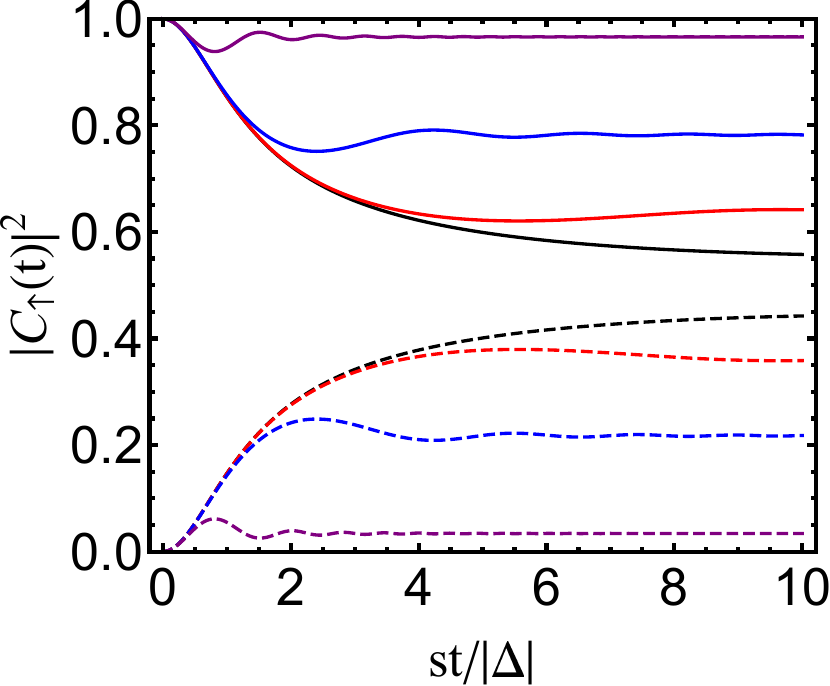}
\caption{Time evolution of the probability $|C_\uparrow(t)|^2$ to be on the upper branch of the adiabatic eigenstates of the Landau-Zener Hamiltonian $H_{\rm LZ}$, Eq.~(\ref{eq:LZ}), cf. Fig.~\ref{fig:semi-LandauZener}. The solid and dashed curves are for $\Delta > 0$ and $\Delta<0$, respectively. The solid curves from top down and the dashed curves from bottom up refer to the same values of $\Delta^2/s{}$. In this order, $\Delta^2/s{} =$ 1.5 (purple), 0.25 (blue), 0.05 (red), 0.01 (black).  The sum of the values of $|C_\uparrow|^2$ on the  solid and dashed curves for the same $\Delta^2/s{}$ (i.e., of the same color) add up to 1 for each time. The initial condition is $\phi(0) = |n-1\rangle.$
}
\label{fig:LandauZener}
\end{figure}

Of primary interest is the long time behavior of  $C_{\uparrow,\downarrow}(t)$. It can be obtained from the asymptotic expansion of the parabolic cylinder functions (see Appendix~\ref{app:Landau_Zener}), or directly by solving the Schr\"odinger equation in the WKB approximation,
\begin{align}
C_\downarrow(t) &\approx \alpha_\downarrow e^{i\theta(t)} +\beta_\downarrow e^{-i\theta(t)}(2s{}t^2)^{-1/2},\nonumber\\
C_\uparrow(t) &\approx  \alpha_\uparrow e^{-i\theta(t)}  +\beta_\uparrow e^{i\theta(t)}(2s{}t^2)^{-1/2},
\label{eq:largetime}
\end{align}
Here, $\theta(t)$ is the dynamical phase $\int_0^t dt' \sqrt{\nu^2(t')+\Delta^2}$ associated with the adiabatic solutions in Fig.~\ref{fig:semi-LandauZener}, 
\begin{align}
\label{eq:phase_factor}
 \theta(t) &= \frac{s{}t^2}{2} + \frac{\Delta^2}{2s{}} \log \left(\frac{2s{} t}{|\Delta|}\right) + \frac{\Delta^2}{4s{}}.  
\end{align}
The expressions for the parameters $\alpha_{\uparrow,\downarrow},\beta_{\uparrow,\downarrow}$ in Eq.~(\ref{eq:largetime}) follow from the general solution of the Schr\"odinger equation; the explicit form of $\alpha_{\uparrow,\downarrow}$ is given in Appendix~\ref{app:Landau_Zener}. 

The coefficients $C_{\uparrow,\downarrow}(t)$ approach their asymptotic values $\propto \alpha_{\uparrow,\downarrow}$ as $1/t$ and oscillate as $\exp[\pm i\theta(t)]$. We note that, for  $t\rightarrow \infty$, we have $C_\uparrow \to C_+(t)$ and $C_\downarrow \to C_-(t)$, i.e., Eq.~(\ref{eq:largetime}) directly gives the coefficients $C_{\pm}$ of the expansion of the wave function in the symmetric and antisymmetric combination of functions $|n\pm 1\rangle$.

\begin{figure}[ht]
\includegraphics[width=8truecm]{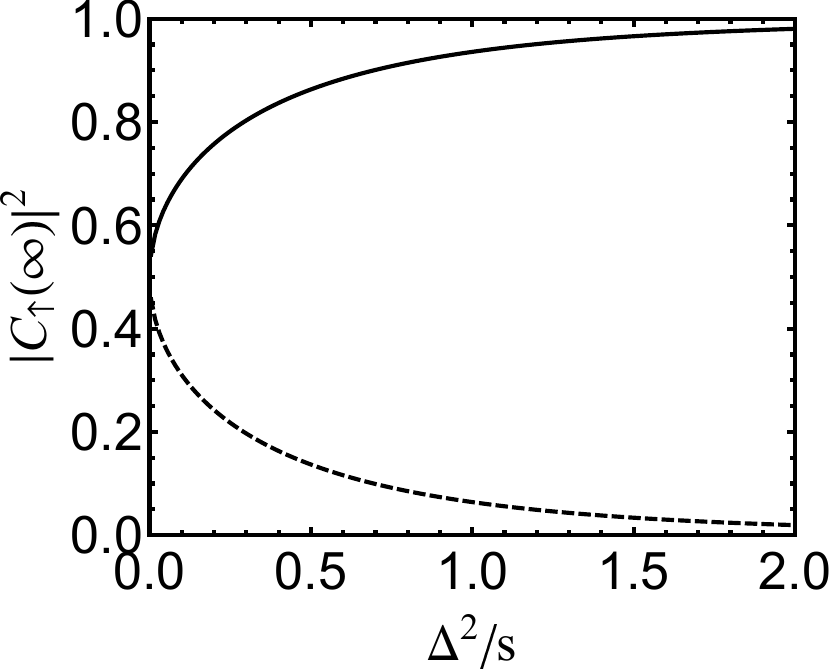}
\caption{The probability $|C_\uparrow(\infty)|^2$ to be on the upper branch of the eigenstates of $H_{\rm LZ}$ in Fig.~\ref{fig:semi-LandauZener} at large time. The solid and dashed lines refer to $\Delta >0$ and $\Delta <0$, respectively. The initial condition is $\phi(0) = |n-1\rangle$.}
\label{fig:nonadiabatic}
\end{figure}

Figure \ref{fig:nonadiabatic} shows the asymptotic value $|C_\uparrow(\infty)|^2 = |\alpha_\uparrow|^2$ as a function of the Landau-Zener parameter $\Delta^2/s{}$.
In the adiabatic limit $\Delta^2/s{}\gg 1$ and for the case $\Delta > 0$,  where the system starts from the upper branch, $\phi(0)=\phi^{\rm ad}_\uparrow (0)$, we have
\begin{align}
\alpha_\uparrow \approx 1- \frac{i}{12}\frac{s}{\Delta^2}, 
\quad \alpha_\downarrow \approx - \frac{i}{4}\frac{ s}{\Delta^2}
\label{eq:adiabatic}
\end{align}
($|\alpha_\uparrow|^2 + |\alpha_\downarrow|^2 = 1$). In distinction from the Landau-Zener problem, where the non-adiabatic transition probability approaches zero exponentially as $\exp(-\pi\Delta^2/s{})$, here it approaches zero as $(\Delta^2/ s)^{-2}$. This special feature is due to the initial condition in the considered problem being set at $t=0$ rather than $t \to -\infty$. 

In the strongly non-adiabatic case, $\Delta^2/s\ll 1$,  if $\phi(0)=|n-1\rangle$,  in the long-time limit the state of the system ultimately approaches an equal superposition of the eigenstates $\phi^{\rm ad}_{\uparrow,\downarrow}$ of $H_{\rm LZ}$: $|\alpha_{\uparrow}| \approx |\alpha_\downarrow| \approx 1/\sqrt{2}.$ This can be seen from Eq.~(\ref{eq:LZ}); see also Appendix~\ref{app:Landau_Zener}. In the case $\Delta = 0$, the states $(|n-1\rangle \pm |n+1\rangle)/\sqrt{2}$ are exact eigenstates for any time $t$. Therefore, $\phi(t)$ will remain in an equal superposition of these two states for any time; note, however, that the states depend on time differently. 

An instructive case is when the oscillator is in the ground state before the driving is applied and the detuning of the driving frequency $\delta\omega_F$ is close to $3V/2$. Here, if the field is ramped up fast, the oscillator will end up in equally populated adiabatic states, which corresponds to two equally populated even interwell states in Fig.~\ref{fig:schematic}.

\section{ Adiabaticity in the presence of dissipation}
\label{sec:dissipation}

Coupling to the environment leads to decoherence of the quasienergy states. It reduces the fidelity of the state preparation. Here we consider the constraint on the dissipation in the case of  state preparation by slowly ramping up the driving field. To achieve high fidelity, one needs to increase the field at a rate larger than the relaxation rate, but smaller than the reciprocal spacing of the relevant RWA energies divided by $\hbar$. For a state $\phi_E$, this means that the decay rate of this state $\Gamma_E$ should be small compared to $\Delta_E$, where $\hbar\Delta_E$ is the instantaneous difference between the quasienergy of the state $\phi_E$ and the nearest state of the same parity. The parity constraint here is the consequence of the fact that the field mixes only the same-parity states.

The RWA level spacing $\hbar \Delta_E$ can be estimated where the driving is weak, $\tilde F\ll V$, or strong, $\tilde F\gg V$. 
For weak driving, the RWA eigenstates are close to the Fock states. From the results of Sec.~\ref{sec:RWA}, $\Delta_E \sim V$ and depends on the ratio $\delta\omega_F/V$, cf. Fig.~\ref{fig:zerodrive}. At strong driving, $\hbar\Delta_E$ is given by the spacing of the intrawell energy levels of the Hamiltonian $H_{\rm RWA}(Q,P)$; see Fig.~\ref{fig:schematic}. It is determined by the frequency $\omega_{\rm min}$ of oscillations about the minima of $H_{\rm RWA}(Q,P)$, which gives $\Delta_E \approx  2[(\delta\omega_F+\tilde F)\tilde F]^{1/2}$; see Appendix \ref{App:semiclassical}.

We illustrate the effect of dissipation using the well-known model \cite{Mandel1995} where the kinetics in the rotating frame is described by the Markov master equation for the density matrix $\rho$ of the form
\begin{align}
\partial_t{\rho}&=i\hbar^{-1}[\rho,H_{\rm RWA}] - \hat\Gamma \rho,  \nonumber \\
\hat\Gamma\rho &= \Gamma(\hat a^\dagger  \hat a \rho - 2\hat a \rho \hat a^\dagger  +\rho \hat a^\dagger  \hat a).
\label{eq:master_equation}
\end{align}
Here, $\Gamma$ is the oscillator relaxation rate and we assume that the temperature of the environment is sufficiently low, $k_BT\ll \hbar\omega_0$.

The decay rate $\Gamma_E$ of an RWA eigenstate $\phi_E$ can be estimated as the decay rate of the diagonal matrix element of the density matrix $\langle \phi_E |\rho|\phi_E\rangle$. Assuming that the system is in state $\phi_E$, i.e., $\rho = |\phi_E\rangle \langle \phi_E|$, and taking into account that the matrix elements of the ladder operators on the states of the same parity are zero, we find from Eq.~(\ref{eq:master_equation})  $\Gamma_E = 2\Gamma \langle \phi_E|a^\dagger  a|\phi_E\rangle$. At weak driving, $\Gamma_E\sim\Gamma$. At strong driving $\Gamma_E$ is determined by the rate of transitions between the intrawell states of $H_{\rm RWA}$ \cite{Marthaler2006},  $\Gamma_E \sim \Gamma \tilde F/V$.

\begin{figure}[ht]
\vspace*{0.1in}
\includegraphics[width = 8truecm]{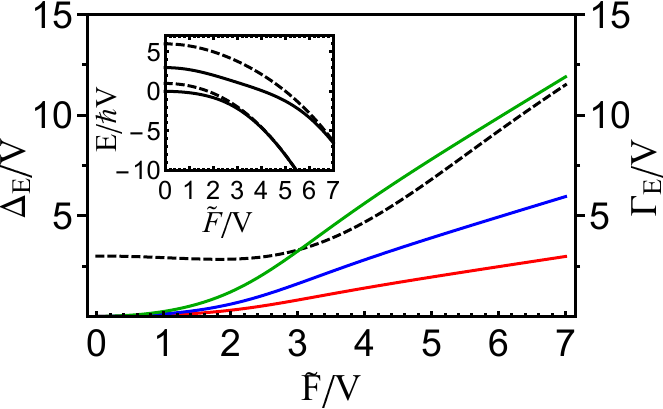}
\caption{Solid lines from top down: the instantaneous decay rate $\Gamma_E= 2\Gamma \langle \phi_E|a^\dagger  a|\phi_E\rangle$ of the state $\phi_E$ for $\Gamma/V =$ 2 (green), 1 (blue) and 0.5 (red).  Dashed line: the instantaneous level spacing $\hbar\Delta_E$ between the state $\phi_E$ and the nearest state of the same parity. The scaled detuning is $\delta\omega_F/V = 0$. The state $\phi_E$ is chosen to be the lowest RWA state, $\phi_E = |0\rangle$ for  $F = 0$. The inset shows the evolution of the RWA spectrum with increasing $F$; the solid and dashed lines refer to the two lowest even- and odd-parity states, respectively.}
\label{fig:quasienergy_decay}
\end{figure} 

From the above estimates, the adiabaticity condition $\Gamma_E \ll \Delta_E$ requires that $\Gamma \ll V,|\delta\omega_F|$ for weak driving and $\Gamma \ll V$ for strong driving.  
Fig.~\ref{fig:quasienergy_decay} illustrates the evolution of $\Delta_E$ and $\Gamma_E$ of an RWA  eigenstate $\phi_E$ with the varying driving amplitude $F$. For the case shown in the figure, the state $\phi_E$ has the lowest RWA eigenenergy.  At large $\tilde F/V$, both $\Delta_E$ and $\Gamma_E$ increase linearly with $F$ as we expect from the analysis above. The slope of $\Gamma_E$ as a function of $\tilde F$ increases as $\Gamma/V$ increases. It coincides with the slope of $\Delta_E$ for $\Gamma/V \approx 2$ as shown by the green curve. For the condition $\Gamma_E \ll \Delta_E$ to be satisfied for any $F$, one needs to have $\Gamma /V \ll 2$. For $\Gamma / V \gtrsim 2$, in the considered case $\Gamma_E$ and $\Delta_E$ as a function of $\tilde F$ can cross each other.

\section{Transient radiation from quasienergy states}

Decay of a parametrically driven oscillator is accompanied by emission of excitations into the surrounding medium. The most familiar picture is decay of optical/microwave cavity modes into propagating electromagnetic waves. Detection of the radiation from the cavity provides a way of characterizing the cavity modes. Radiation from the modes in a non-steady state, such as a quasienergy state, is transient. After a time of the order of the mode relaxation time, the system relaxes to a steady state, the radiation becomes steady and does not depend on the quasienergy state the system had been staying in. To identify a quasienergy state from the radiation, one needs to collect the transient radiation.

\begin{widetext}
\onecolumngrid
\begin{figure}[ht]
\includegraphics[width = 15truecm]{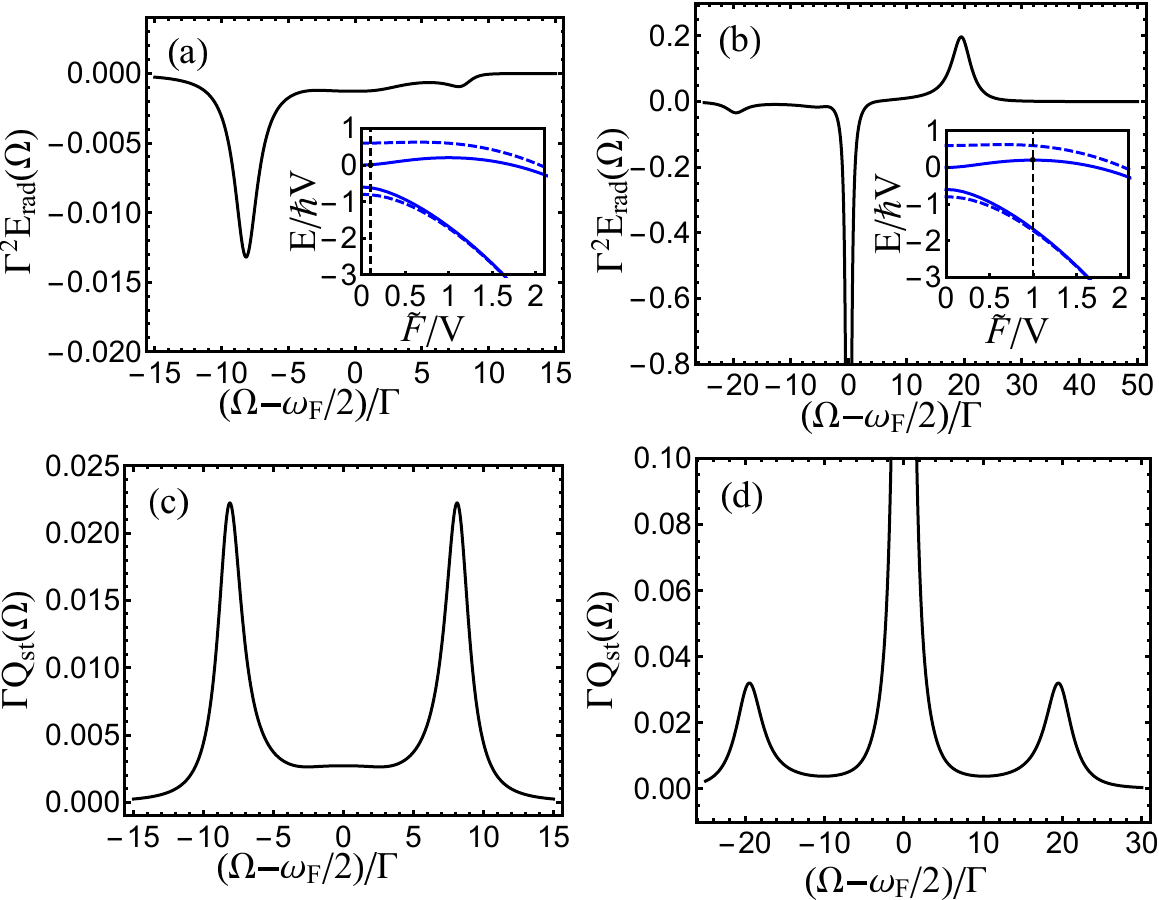}
\caption{The transient and steady state spectra of radiation emitted by a parametrically driven oscillator. The scaled detuning is $\delta\omega_F/V = 1.8$, the scaled decay rate is $\Gamma/V = 0.1$. Panels (a) and (b): The transient spectrum $E_{\rm rad}(\Omega)$, Eq.~(\ref{eq:transient_energy}),  for the second lowest even RWA eigenstate, which can be prepared adiabatically from the oscillator ground state $|0\rangle$ by ramping up the field to $\tilde F/V = 0.1$ in (a) and to $\tilde F/V = 1$ in (b). The insets show the dependence of the RWA energy levels on $F$; the adjacent (blue) dashed and solid lines refer to odd and even states, respectively, whereas the vertical (black) dashed lines indicate the above driving amplitudes. Panels (c) and (d): The steady state power spectrum with the same parameters as in (a) and (b), respectively.}
\label{fig:transient_radiation}
\end{figure} 
\end{widetext}

We model the radiation field by a set of oscillators enumerated by subscript $k$,  with quasi-continuous frequencies $\omega_k$ and with Hamiltonian $H_{\rm rad} = \sum_k \hbar\omega_k b^\dagger_k b_k$. We assume that the coupling of the considered oscillator to this field is bilinear in the ladder operators of the oscillator and the radiation, $H_i = \sum_k \xi_k (b_k+b_k^\dagger) (a+a^\dagger )$, where  $\xi_k$ are the coupling parameters. The total Hamiltonian is $H_{\rm total} = H_0 + H_{\rm rad} + H_{i}$. Operator $H_0$ is the Hamiltonian of the oscillator and the non-radiative thermal reservoir to which the oscillator is coupled. We assume that this reservoir and the radiation field are at the same temperature, which we assume to be sufficiently low, $k_BT\ll \hbar\omega_0$.  The coupling to the reservoir leads to relaxation of the oscillator with typical relaxation rate $\Gamma$, cf. Eq.~(\ref{eq:master_equation}).\footnote{The oscillator is characterized also by a much longer rate, which is related to the dissipation-induced transitions between the wells in Fig.~1.}

If the coupling to the radiation field is weak, it can be considered as a perturbation to the non-radiative dynamics. The power of the radiation emitted into a spectral range $d\Omega$ around frequency $\Omega$ is given by the change of the energy of the radiation field in this interval per unit time $W(\Omega,t)d\Omega =  \frac{d}{dt} \sum_k \delta(\omega_k-\Omega) d\Omega\langle  \hbar \omega_k b_k^\dagger b_k \rangle$. To the lowest order in the coupling strength $\xi_k$, we have in the resonant region where $\Omega$ is close to $\omega_F/2$ \cite{Dykman1975} 
\begin{align}
&W(\Omega,t) = Q[\Omega,t-t_0,\rho_0(t_0)]\Omega\xi^2(\Omega),  \nonumber \\  
&Q[\Omega,t-t_0,\rho_0(t_0)]  = 2\Re \int_{t_0}^t d{}t' e^{i(\Omega-\omega_F/2)(t-{t'})} \nonumber \\
& \times \Tr[ a^\dagger ({}t'-t_0) a(t-t_0)\rho_0(t_0)],
\label{eq:transient_power}
\end{align}
where $\xi^2(\Omega) = \hbar^{-1}\sum_k|\xi_k|^2\delta(\Omega-\omega_k)$.

In deriving Eq.~(\ref{eq:transient_power}) we assumed that the coupling to the radiation is switched on at time $t_0$;  $\rho_0(t_0)$ is the density matrix of the oscillator and the non-radiative environment. Equation~(\ref{eq:transient_power}) is written in the rotating frame used above to find the quasienergy states of the oscillator, with the time counted off from $t_0$.

The two-time correlation function in Eq.~(\ref{eq:transient_power}) can be found by solving the quantum kinetic equation. As an initial condition to this equation we choose the density matrix $\rho_0(t_0)$ in the form of a product of the oscillator density matrix $\rho(t_0)$ and the density matrix of the non-radiative environment in thermal equilibrium.   Such choice is justified, since a weak coupling to the dissipative (non-radiative) environment allows preparing the oscillator in a certain state at time $t_0$ given that the preparation time is short compared to the relaxation time. The following evolution on the time scale, which largely exceeds both $\omega_F^{-1}$ and the time it took to prepare the state, can be described by assuming that at $t_0$ there is switched on not only the coupling to the radiation field, but also the stronger (but still weak) coupling to the non-radiative environment. Corrections to the dynamics due to the switching are well-understood, they are small in the considered case \cite{Dykman1975}. 

The time evolution of the oscillator density matrix in the rotating frame is then often described by Eq.~(\ref{eq:master_equation}). To study transient radiation, we set $\rho(t_0)=|\phi_E\rangle \langle \phi_E \rangle$, where $\phi_E$ is a RWA eigenstate in which the oscillator is prepared at $t_0$.

For not very strong driving, $\tilde F \lesssim V$, function $\phi_E$ has a contribution of only a few Fock states. Respectively,  the oscillator will radiate only a few photons as it comes to the stationary state. Then, rather than measuring the radiation power $W(\Omega,t)$ it is more feasible to measure the total energy emitted over the transient time. The observation time should exceed the relaxation time to enable sufficient spectral resolution. 

The energy of the transient radiation has to be separated from the energy that the oscillator emits in the stationary state. This can be done by noting that the latter energy is proportional to the observation time. The spectral power density (power per unit frequency)  in the stationary regime is given by Eq.~(\ref{eq:transient_power}) written for $t\to \infty$ \cite{Dykman2012}. Therefore one can define the transient radiation spectral density as the integral over time of the difference of the emitted power (\ref{eq:transient_power}) and the power emitted in the stationary regime. Writing this spectral density as $\Omega\xi^2(\Omega)E_{\rm rad}(\Omega)$, we obtain 
\begin{equation}
E_{\rm rad}(\Omega) = \int_{t_0}^\infty dt Q[\Omega,t-t_0,\rho_0(t_0)-\rho_{\rm st}].
\label{eq:transient_energy}
\end{equation}
Here, $\rho_{\rm st}$ is the stationary density matrix of the driven oscillator and the non-radiative environment.

The spectral density $E_{\rm rad}(\Omega)$ is given by the difference between the irradiated energy and the energy that would be irradiated into the same spectral interval if the system were stationary. This difference is accumulated over a sufficiently long time that largely exceeds the relaxation time. By construction, it can be positive or negative.
%
%

As the oscillator decays from the initial state $\phi_E$, it emits radiation at frequencies  $\omega_F/2+ (E-E')/\hbar$, where $E'$ is the RWA energy of a state $\phi_{E'}$ into which the oscillator can make a dipolar transition from $\phi_E$. In contrast, in the stationary state, the oscillator generally can be found in the both states $\phi_E, \phi_{E'}$, with different probabilities. Depending on these probabilities, it radiates at the both frequencies $\omega_F/2\pm (E-E')/\hbar$  generally with different intensities. As a result, in the  spectrum $E_{\rm rad}(\Omega)$ one may expect a peak or a dip at $\omega_F/2+ (E-E')/\hbar$, but only a dip at $\omega_F/2 - (E-E')/\hbar$.

Figures~\ref{fig:transient_radiation} (a) and (b) show the spectrum $E_{\rm rad}(\Omega)$ when the oscillator is initially in the RWA eigenstate $\phi_E$ prepared from the vacuum $|0 \rangle$ by adiabatically ramping up the driving field. The driving frequency is chosen so that $\phi_E$ has the second lowest RWA energy among even states; see the insets. The transient radiation is dominated by transitions from the state $\phi_E$ to the lowest odd state $\phi_{E'}$. In this case, for a strong driving field $E-E'\approx 2\hbar[\tilde F(\delta\omega_F + \tilde F)]^{1/2}$ corresponds to the spacing between the two lowest intrawell states of $H_{\rm RWA}$ in Fig.~\ref{fig:schematic}. For weak driving, $E-E'\approx \hbar (\delta\omega_F - V)$; the frequency $\omega_F/2 - (E-E')/\hbar = \omega_0 + V $ is the frequency of the transition from the first excited state to the ground state of the undriven oscillator. Figures~\ref{fig:transient_radiation} (a) and (b) refer not to these limiting cases but to the intermediate field strengths.

As expected, the spectrum $E_{\rm rad}(\Omega)$ displays a peak at $\omega_F/2 + (E - E')/\hbar$ for relatively strong driving and a small dip at this frequency for weak driving. It also displays a characteristic pronounced dip at $\omega_F/2 -( E - E')/\hbar$ in the both cases. 
In addition, for strong driving, the spectrum has a negative narrow peak at $\omega_F/2$ due to the interwell transitions \cite{Dykman2012}. 
For a comparison, Figs.~\ref{fig:transient_radiation} (c) and (d) show the steady-state radiation power spectrum $Q_{\rm st}(\Omega) = Q(\Omega,\infty,\rho_{\rm st})$ for the same parameters as in Figure~\ref{fig:transient_radiation}(a) and (b), respectively. 

\section{Conclusions}
\label{sec:conclusions}

We have studied preparation of quasienergy states of a nonlinear oscillator. We found that various states can be prepared with high accuracy in a finite time by simply linearly increasing in time the amplitude of the parametric driving. The driving frequency $\omega_F$ was chosen to be close to twice the oscillator eigenfrequency $\omega_0$, so that strong excitation of the oscillator could be achieved for a comparatively weak driving field. The prepared state sensitively depends on the interrelation between the detuning of the driving frequency $\delta\omega_F=\omega_F/2-\omega_0$ and the nonequidistance $V$ of the oscillator energy levels due to the nonlinearity (in frequency units). 

An important factor for the state preparation is that the quasienergy states are either even or  odd with respect to inversion in the phase space. The states of different parity are not coupled by the driving. This allows one to prepare on demand an arbitrary even quasienergy state just by slowly ramping up the driving, if the oscillator is initially in the ground state. The resulting states have very different structures in phase space, as evidenced by the Wigner tomography. A similar analysis shows that an arbitrary odd state can be prepared, if  initially the oscillator is in the first excited state.

A remarkable feature of the system related to its symmetry is that the oscillator energy levels calculated in the rotating wave approximation do not cross or anti-cross with the increasing  driving amplitude. Rather the neighboring RWA energy levels of even and odd states approach each other pairwise. At the same time, the levels of the opposite-parity states can cross with varying $\delta\omega_F$. This crossing does not lead to crossing of the quasienergy levels. Where the RWA energy levels cross, the quasienergy levels are separated by $\hbar\omega_F/2$.

It is also important for the state preparation that, in the limit of zero driving, the RWA energy spectrum can simultaneously have several double-degenerate levels. Such degeneracy corresponds to either a multi-photon or a subharmonic resonance. By tuning the driving frequency, one can bring the RWA energy levels closer or further away from the pairwise degeneracy. 

The degeneracy of same-parity states provides an effective way of preparing superpositions of quasienergy states. It is based on non-adiabatic transitions induced by the increasing driving amplitude. The field leads to the state mixing that depends on how fast it is increased. The problem differs from the standard Landau-Zener problem, since the initial state is close to degeneracy and the field is ramped up in a finite time. As a result, for a linearly increasing  field, the probability of the non-adiabatic transition falls off as a power law, rather than exponentially, with the Landau-Zener parameter $\Delta^2/s$, where $\Delta$ is the level spacing and $s$ is the ramp-up speed. 

Dissipation due to the coupling to a thermal reservoir reduces the fidelity of the state preparation. However, away from the level degeneracy, the effect of the dissipation is small, if the oscillator nonlinearity parameter $V$ exceeds the decay rate $\Gamma$. Then one can ramp up the driving at a rate that is much smaller than the quasienergy level spacing, yet much larger than the decay rate. Fluctuations of the system parameters and of the driving power can also reduce the fidelity. Their effect is small if their bandwidth is small compared to $V$ or if they are sufficiently weak, so that their effect does not accumulate over the duration of the state preparation.

Because of dissipation, the parametric oscillator prepared in a given quasienergy state will ultimately come to a stationary state. Our results show that the prepared state can be characterized by studying the transient radiation of the oscillator. This method is complimentary to the commonly used Wigner tomography. It can be particularly useful for investigating quasienergy states of cavity modes in microwave cavities, the area of much current interest. The above analysis suggests a simple way of preparing various quasienergy states in such cavities as well as in other systems, for example, Josephson junctions, that can be modeled by nonlinear quantum parametric oscillators.

\section{Acknowledgements}
This work was supported in part by the National Science Foundation (Grant No. DMR-1514591); YZ was also partly supported by the U.S. Army Research Office (W911NF1410011) and by the National Science Foundation (DMR-1609326).

\appendix
\section{Fourier series for quasienergy states}
\label{app:Floquet}

The eigenvalue problem for the periodic part $u_\ep(t)$ of the Floquet wave function defined in Eq.~(\ref{eq:Floquetdef}) reads
\begin{equation}
\ep u_\ep(t) = (H(t)-i\hbar\partial_t)u_\ep(t).
\label{eq:Floquetequation}
\end{equation}
Since $u_\ep(t)$ and $H(t)$ are both periodic in time, it is convenient to expand them in Fourier series. It is also convenient to write $u_\ep(t)$  in the basis of the Fock states $|n\rangle$ of the harmonic oscillator with frequency $\omega_0$. Then
$u_\ep(t) = \sum_{k,n}u_{k,n}\exp(-ik\omega_Ft)|n\rangle$ and Eq.~(\ref{eq:Floquetequation}) takes the form of the standard eigenvalue problem
\begin{align}
\ep u_{k,n} =&\sum_{k',n'} M^{k',n'}_{k,n} u_{k',n'},  \nonumber\\
M^{k',n'}_{k,n} = & (\mathcal E_n-k\hbar\omega_F ) \delta_{k,k'}\delta_{n,n'}\nonumber\\
& + \frac{1}{4}F q^2_{nn'}(\delta_{k',k+1}+\delta_{k',k-1})
\label{eq:FloquetFourier}
\end{align}
where $q^2_{nm} = \langle n| q^2| m\rangle$ and $\mathcal E_n$ is the $n$th energy level of the Duffing oscillator in the absence of driving; to the leading order in the nonlinearity $\mathcal E_n= \hbar[ \omega_0 n + V(n^2+n)/2]$. The sum runs over $k = 0,\pm1,\pm2, ...$ and $ n = 0,1,2,...$.  

The matrix elements $q^2_{nm}$ are nonzero for $n=m$ and $n=m\pm 2$. Therefore the driving term $\propto F$ couples $ u_{k,n}$ to $ u_{k\pm1,n\pm2},  u_{k\pm1,n}$. However, only the coupling to $ u_{k+1,n+2}$ and $u_{k-1,n-2}$ is resonant, since the diagonal elements of matrix $\hat M$ for such $u$ are close; for example, $ (\mathcal E_n-k\hbar\omega_F ) - 
 [\mathcal E_{n+2} - (k+1)\hbar\omega_F ] = 2\hbar\delta\omega_F-\hbar V(2n+3)$ is small compared to $\hbar\omega_F$.
Therefore, one can limit the analysis to a set $G_{k,n}$ of the variables $u_{k',n'}$ resonantly coupled to $u_{k,n}$. It has the form  $G_{k,n}=\{ u_{k+k',n+2k'}, k'\in \mathbb{Z} \text{ and } k'\geq -n/2\}$. This is the rotating wave approximation in the Floquet formulation (\ref{eq:Floquetequation}).

The sets $G_{k,n}$ with different $k$ but the same $n$ are equivalent: indeed, changing $k\to k_1$ corresponds to changing $\ep \to \ep + (k-k_1)\hbar\omega_F$ in Eq.~(\ref{eq:FloquetFourier}). Since $\ep$ is defined modulo $\hbar\omega_F$, such change makes no difference. We can then simplify $G_{k,n}$ as follows. Consider first even $n$, i.e., $n=2n'$, and set $k=n'$, 
\begin{align}
&G_{k,2n'}\equiv G_{n',2n'} \nonumber \\ 
&= \{  u_{n'+k',2n'+2k'},k'=-n',-n'+1,...\}=G_{0,0}.
\label{eq:evenG}
\end{align}
In the last equation, we simply redefined $k'$ to absorb $n'$ in the new definition.

Similarly, for odd $n$, where $n=2n'+1$,
\begin{equation}
G_{k,2n'+1}\equiv G_{0,1}.
\label{eq:oddG}
\end{equation}

The simplification described by Eqs.~(\ref{eq:evenG}) and (\ref{eq:oddG}) allows one to reduce  Eq.~(\ref{eq:FloquetFourier}) to two sets of equations, 
\begin{align}
&(\ep-\mathcal E_{2k}+k\hbar \omega_F)  u_{k,2k} \nonumber\\
&=\tilde F\left[k(2k-1) u_{k-1,2k-2}+(k+1)(2k+1)  u_{k+1,2k+2}\right], \nonumber\\
&(\ep -\mathcal E_{2k+1}+k\hbar \omega_F) u_{k,2k+1}\nonumber\\
&= \tilde F\left[k(2k+1) u_{k-1,2k-1}+(k+1)(2k+3) u_{k+1,2k+3}\right] 
\label{eq:FloquetFouriersimplify}
\end{align}
where $\tilde F = F/4\omega_0 \approx F/2\omega_F$ and we used the explicit form of the matrix elements $\langle n|q^2|n\pm 2\rangle$.

Equation~(\ref{eq:FloquetFouriersimplify}) coincides with the RWA  Schr\"odinger equation $E\phi_E = H_{\rm RWA}\phi_E$ if one writes $\phi_E$ in the basis of the Fock states and replaces $\ep$ with $E$ using Eq.~(\ref{eq:RWAFloquetrelation}).

\section{Semiclassical analysis of RWA Hamiltonian}
\label{App:semiclassical}

For completeness, here we present, following \cite{Marthaler2006}, the description of the scaled RWA Hamiltonian function $g(Q,P)$ for large driving. For $\mu < -1$, function $g$ has one minimum at $(Q,P) = (0,0)$. For $ -1 <\mu < 1$, the minimum at (0,0) becomes a saddle point and there appears two minima located at $(Q,P) = (\pm Q_0,0), Q_0 = \sqrt{\mu+1}$. For $\mu>1$, the saddle point at $(0,0)$ becomes a minimum again and there appear two saddle points at $(Q,P) = (0,\pm\sqrt{\mu-1})$. 

Of primary interest in this paper is the regime $0 \leq \mu < 1$ where the quasinergy spectrum can display degeneracy and RB degeneracy. We expand $g$ about the minimum at  $(Q_0,0)$ to second order in $Q-Q_0$ and $P$,
\begin{equation}
g \approx (\mu+1)(Q-Q_0)^2 + P^2  + g_{\rm min},
\end{equation}
where $g_{\rm min} = -(\mu+1)^2/4$. 

Introducing ladder operators $b,b^\dagger$ defined as 
\[
Q-Q_0 = \sqrt{\frac{\lambda}{2}}(\mu+1)^{-1/4}(b^\dagger +b) , \]
\[
P = i\sqrt{\frac{\lambda}{2}}(\mu+1)^{1/4}(b^\dagger -b) \]
($[b,b^\dagger]=1$), we write the Hamiltonian $g(Q,-i\lambda\partial_Q)$ for low-lying intrawell eigenstates  in the form
\begin{align}
&g \approx \lambda\omega_{\rm min}(b^\dagger b + 1/2) + g_{\rm min}, \nonumber \\
&\omega_{\rm min} = 2\sqrt{\mu+1}.
\end{align}
The eigenstates of operator $b^\dagger b$ give the intra-well states used in the main text.

\section{Non-adiabatic transition amplitude}
\label{app:Landau_Zener}

The equations for $C_\pm(t)$ in Sec.~\ref{sec:non-adiabatic} can be rescaled to the form of Weber differential equation,
\begin{align}
\frac{d^2 C_\pm}{dz_\pm^2} + \left [-\frac{z_\pm^2}{4} \mp ip +\frac{1}{2}\right ]C_\pm =0, \nonumber \\
p =\Delta^2/2s{},\qquad  z_\pm = \sqrt{2s{}} e^{\pm i\pi/4}t.
\label{eq:Weber}
\end{align}

The general solution to this equation is a linear combination of two parabolic cylinder functions \cite{Whittaker1990},
\begin{align}
C_\pm(z) = A_{\pm} D_{\pm ip -1}(\mp iz_\pm) + B_{\pm} D_{\mp ip}(z_\pm).
\label{eq:parabolic}
\end{align} 
Coefficients $A_{\pm},B_{\pm}$ can be found from the initial values of $C_\pm(0)$ with account taken of the relation  $i\hbar\dot C_{\pm} (0) = \Delta C_{\mp}(0)$. 

Using the asymptotic expansion 
$D_q(z) \approx \exp(-z^2/4)z^q$ for $|z| \rightarrow \infty, |\rm{arg}\, z| < \frac{3}{4}\pi$,
we find to the first order in $1/t$ 
\begin{align}
C_\pm(t) &\approx B_\pm \alpha_\pm  e^{\mp i\theta(t)} + A_\pm\alpha^*_\pm  e^{\pm i\theta(t)+i\pi/4}(2st^2)^{-1/2}, \nonumber \\
 \alpha_+ &=  \exp\left[\frac{p\pi}{4}  - \frac{ip}{2}(\log p-1)  \right]=  \alpha_-^*,
\end{align}
where $\theta(t)$ is given by Eq.~(\ref{eq:phase_factor}). 

For $|\Delta| \ll \nu(t)$ we have $C_\uparrow \approx C_+ + (\Delta/2\nu)C_-$ and   $C_\downarrow \approx C_- - (\Delta/2\nu)C_+$. One can then immediately find the coefficients $\alpha_{\uparrow,\downarrow},\beta_{\uparrow,\downarrow}$ in Eq.~(\ref{eq:largetime}). In particular, $\alpha_{\uparrow} = B_+\alpha_+, \alpha_{\downarrow} = B_-\alpha_-.$

Of primary interest to us is the limiting value $C_{\uparrow,\downarrow}(\infty) \propto \alpha_{\uparrow,\downarrow}$. For the considered initial condition $C_+(0)=C_-(0)=1/\sqrt{2}$, we find that 
\begin{align}
\alpha_{\uparrow,\downarrow} &= \Lambda_\pm \left[ \sqrt{p}\,\Gamma\left(\mp \frac{ip}{2}\right) + \sgn(\Delta) (\pm 1+ i)\Gamma\left(\frac{1\mp ip}{2}\right) \right], \nonumber \\
\Lambda_+ &= \Lambda_-^* = (2p/e)^{-ip/2}(e^{3\pi p/4}- e^{-5\pi p/4})\nonumber\\
&\qquad \times\sqrt{p}\, \Gamma(ip) /4\sqrt{2}\pi. 
\label{eq:alpha}
\end{align}
where the upper sign refers to $\alpha_\uparrow$ and the lower sign refers to $\alpha_\downarrow$; $\Gamma(x)$ is the gamma function.

The expressions for $\alpha_{\uparrow,\downarrow}$ in the adiabatic limit $p\rightarrow \infty$ can be obtained from Eqs.~(\ref{eq:alpha}) using the asymptotic form of the gamma function $\Gamma(z)$ for $|z|\to \infty$, cf.  \cite{Abramowitz1972}. They were used in Eq.~(\ref{eq:adiabatic}). 

\bibliographystyle{apsrev4-1}

\begin{thebibliography}{36}%
\makeatletter
\providecommand \@ifxundefined [1]{%
 \@ifx{#1\undefined}
}%
\providecommand \@ifnum [1]{%
 \ifnum #1\expandafter \@firstoftwo
 \else \expandafter \@secondoftwo
 \fi
}%
\providecommand \@ifx [1]{%
 \ifx #1\expandafter \@firstoftwo
 \else \expandafter \@secondoftwo
 \fi
}%
\providecommand \natexlab [1]{#1}%
\providecommand \enquote  [1]{``#1''}%
\providecommand \bibnamefont  [1]{#1}%
\providecommand \bibfnamefont [1]{#1}%
\providecommand \citenamefont [1]{#1}%
\providecommand \href@noop [0]{\@secondoftwo}%
\providecommand \href [0]{\begingroup \@sanitize@url \@href}%
\providecommand \@href[1]{\@@startlink{#1}\@@href}%
\providecommand \@@href[1]{\endgroup#1\@@endlink}%
\providecommand \@sanitize@url [0]{\catcode `\\12\catcode `\$12\catcode
  `\&12\catcode `\#12\catcode `\^12\catcode `\_12\catcode `\%12\relax}%
\providecommand \@@startlink[1]{}%
\providecommand \@@endlink[0]{}%
\providecommand \url  [0]{\begingroup\@sanitize@url \@url }%
\providecommand \@url [1]{\endgroup\@href {#1}{\urlprefix }}%
\providecommand \urlprefix  [0]{URL }%
\providecommand \Eprint [0]{\href }%
\providecommand \doibase [0]{http://dx.doi.org/}%
\providecommand \selectlanguage [0]{\@gobble}%
\providecommand \bibinfo  [0]{\@secondoftwo}%
\providecommand \bibfield  [0]{\@secondoftwo}%
\providecommand \translation [1]{[#1]}%
\providecommand \BibitemOpen [0]{}%
\providecommand \bibitemStop [0]{}%
\providecommand \bibitemNoStop [0]{.\EOS\space}%
\providecommand \EOS [0]{\spacefactor3000\relax}%
\providecommand \BibitemShut  [1]{\csname bibitem#1\endcsname}%
\let\auto@bib@innerbib\@empty
\bibitem [{\citenamefont {Shirley}(1965)}]{Shirley1965}%
  \BibitemOpen
  \bibfield  {author} {\bibinfo {author} {\bibfnamefont {J.~H.}\ \bibnamefont
  {Shirley}},\ }\href {\doibase 10.1103/PhysRev.138.B979} {\bibfield  {journal}
  {\bibinfo  {journal} {Phys. Rev.}\ }\textbf {\bibinfo {volume} {138}},\
  \bibinfo {pages} {B979} (\bibinfo {year} {1965})}\BibitemShut {NoStop}%
\bibitem [{\citenamefont {Zel'dovich}(1967)}]{Zeldovich1967}%
  \BibitemOpen
  \bibfield  {author} {\bibinfo {author} {\bibfnamefont {Y.~B.}\ \bibnamefont
  {Zel'dovich}},\ }\href@noop {} {\bibfield  {journal} {\bibinfo  {journal}
  {JETP}\ }\textbf {\bibinfo {volume} {24}},\ \bibinfo {pages} {1006} (\bibinfo
  {year} {1967})}\BibitemShut {NoStop}%
\bibitem [{\citenamefont {Ritus}(1967)}]{Ritus1967}%
  \BibitemOpen
  \bibfield  {author} {\bibinfo {author} {\bibfnamefont {V.~I.}\ \bibnamefont
  {Ritus}},\ }\href@noop {} {\bibfield  {journal} {\bibinfo  {journal} {JETP}\
  }\textbf {\bibinfo {volume} {24}},\ \bibinfo {pages} {1041} (\bibinfo {year}
  {1967})}\BibitemShut {NoStop}%
\bibitem [{\citenamefont {Sambe}(1973)}]{Sambe1973}%
  \BibitemOpen
  \bibfield  {author} {\bibinfo {author} {\bibfnamefont {H.}~\bibnamefont
  {Sambe}},\ }\href {\doibase 10.1103/PhysRevA.7.2203} {\bibfield  {journal}
  {\bibinfo  {journal} {Phys. Rev. A}\ }\textbf {\bibinfo {volume} {7}},\
  \bibinfo {pages} {2203} (\bibinfo {year} {1973})}\BibitemShut {NoStop}%
\bibitem [{\citenamefont {Kitagawa}\ \emph {et~al.}(2010)\citenamefont
  {Kitagawa}, \citenamefont {Berg}, \citenamefont {Rudner},\ and\ \citenamefont
  {Demler}}]{Kitagawa2010}%
  \BibitemOpen
  \bibfield  {author} {\bibinfo {author} {\bibfnamefont {T.}~\bibnamefont
  {Kitagawa}}, \bibinfo {author} {\bibfnamefont {E.}~\bibnamefont {Berg}},
  \bibinfo {author} {\bibfnamefont {M.}~\bibnamefont {Rudner}}, \ and\ \bibinfo
  {author} {\bibfnamefont {E.}~\bibnamefont {Demler}},\ }\href
  {http://link.aps.org/doi/10.1103/PhysRevB.82.235114} {\bibfield  {journal}
  {\bibinfo  {journal} {Phys. Rev. B}\ }\textbf {\bibinfo {volume} {82}},\
  \bibinfo {pages} {235114} (\bibinfo {year} {2010})}\BibitemShut {NoStop}%
\bibitem [{\citenamefont {Lindner}\ \emph {et~al.}(2011)\citenamefont
  {Lindner}, \citenamefont {Refael},\ and\ \citenamefont
  {Galitski}}]{Lindner2011}%
  \BibitemOpen
  \bibfield  {author} {\bibinfo {author} {\bibfnamefont {N.~H.}\ \bibnamefont
  {Lindner}}, \bibinfo {author} {\bibfnamefont {G.}~\bibnamefont {Refael}}, \
  and\ \bibinfo {author} {\bibfnamefont {V.}~\bibnamefont {Galitski}},\ }\href
  {http://dx.doi.org/10.1038/nphys1926} {\bibfield  {journal} {\bibinfo
  {journal} {Nat Phys}\ }\textbf {\bibinfo {volume} {7}},\ \bibinfo {pages}
  {490} (\bibinfo {year} {2011})}\BibitemShut {NoStop}%
\bibitem [{\citenamefont {Goldman}\ \emph {et~al.}(2014)\citenamefont
  {Goldman}, \citenamefont {Juzeliunas}, \citenamefont {Ohberg},\ and\
  \citenamefont {Spielman}}]{Goldman2014}%
  \BibitemOpen
  \bibfield  {author} {\bibinfo {author} {\bibfnamefont {N.}~\bibnamefont
  {Goldman}}, \bibinfo {author} {\bibfnamefont {G.}~\bibnamefont {Juzeliunas}},
  \bibinfo {author} {\bibfnamefont {P.}~\bibnamefont {Ohberg}}, \ and\ \bibinfo
  {author} {\bibfnamefont {I.~B.}\ \bibnamefont {Spielman}},\ }\href {\doibase
  10.1088/0034-4885/77/12/126401} {\bibfield  {journal} {\bibinfo  {journal}
  {Reports On Progress In Physics}\ }\textbf {\bibinfo {volume} {77}},\
  \bibinfo {pages} {126401} (\bibinfo {year} {2014})}\BibitemShut {NoStop}%
\bibitem [{\citenamefont {Bukov}\ \emph {et~al.}(2015)\citenamefont {Bukov},
  \citenamefont {D'Alessio},\ and\ \citenamefont {Polkovnikov}}]{Bukov2015}%
  \BibitemOpen
  \bibfield  {author} {\bibinfo {author} {\bibfnamefont {M.}~\bibnamefont
  {Bukov}}, \bibinfo {author} {\bibfnamefont {L.}~\bibnamefont {D'Alessio}}, \
  and\ \bibinfo {author} {\bibfnamefont {A.}~\bibnamefont {Polkovnikov}},\
  }\href {http://dx.doi.org/10.1080/00018732.2015.1055918} {\bibfield
  {journal} {\bibinfo  {journal} {Adv. Phys.}\ }\textbf {\bibinfo {volume}
  {64}},\ \bibinfo {pages} {139} (\bibinfo {year} {2015})}\BibitemShut
  {NoStop}%
\bibitem [{\citenamefont {Peano}\ \emph {et~al.}(2016)\citenamefont {Peano},
  \citenamefont {Houde}, \citenamefont {Brendel}, \citenamefont {Marquardt},\
  and\ \citenamefont {Clerk}}]{Peano2016}%
  \BibitemOpen
  \bibfield  {author} {\bibinfo {author} {\bibfnamefont {V.}~\bibnamefont
  {Peano}}, \bibinfo {author} {\bibfnamefont {M.}~\bibnamefont {Houde}},
  \bibinfo {author} {\bibfnamefont {C.}~\bibnamefont {Brendel}}, \bibinfo
  {author} {\bibfnamefont {F.}~\bibnamefont {Marquardt}}, \ and\ \bibinfo
  {author} {\bibfnamefont {A.~A.}\ \bibnamefont {Clerk}},\ }\href
  {http://dx.doi.org/10.1038/ncomms10779} {\bibfield  {journal} {\bibinfo
  {journal} {Nat. Comm.}\ }\textbf {\bibinfo {volume} {7}},\ \bibinfo {pages}
  {10779} (\bibinfo {year} {2016})}\BibitemShut {NoStop}%
\bibitem [{\citenamefont {Khemani}\ \emph {et~al.}(2016)\citenamefont
  {Khemani}, \citenamefont {Lazarides}, \citenamefont {Moessner},\ and\
  \citenamefont {Sondhi}}]{Khemani2016}%
  \BibitemOpen
  \bibfield  {author} {\bibinfo {author} {\bibfnamefont {V.}~\bibnamefont
  {Khemani}}, \bibinfo {author} {\bibfnamefont {A.}~\bibnamefont {Lazarides}},
  \bibinfo {author} {\bibfnamefont {R.}~\bibnamefont {Moessner}}, \ and\
  \bibinfo {author} {\bibfnamefont {S.~L.}\ \bibnamefont {Sondhi}},\
  }\href@noop {} {\bibfield  {journal} {\bibinfo  {journal} {Phys. Rev. Lett.}\
  }\textbf {\bibinfo {volume} {116}},\ \bibinfo {pages} {250401} (\bibinfo
  {year} {2016})}\BibitemShut {NoStop}%
\bibitem [{\citenamefont {von Keyserlingk}\ and\ \citenamefont
  {Sondhi}(2016)}]{Keyserlingk2016}%
  \BibitemOpen
  \bibfield  {author} {\bibinfo {author} {\bibfnamefont {C.~W.}\ \bibnamefont
  {von Keyserlingk}}\ and\ \bibinfo {author} {\bibfnamefont {S.~L.}\
  \bibnamefont {Sondhi}},\ }\href {\doibase 10.1103/PhysRevB.93.245146}
  {\bibfield  {journal} {\bibinfo  {journal} {Phys. Rev. B}\ }\textbf {\bibinfo
  {volume} {93}},\ \bibinfo {pages} {245146} (\bibinfo {year}
  {2016})}\BibitemShut {NoStop}%
\bibitem [{\citenamefont {{Khemani}}\ \emph {et~al.}(2016)\citenamefont
  {{Khemani}}, \citenamefont {{von Keyserlingk}},\ and\ \citenamefont
  {{Sondhi}}}]{Khemani2016a}%
  \BibitemOpen
  \bibfield  {author} {\bibinfo {author} {\bibfnamefont {V.}~\bibnamefont
  {{Khemani}}}, \bibinfo {author} {\bibfnamefont {C.~W.}\ \bibnamefont {{von
  Keyserlingk}}}, \ and\ \bibinfo {author} {\bibfnamefont {S.~L.}\ \bibnamefont
  {{Sondhi}}},\ }\href@noop {} {\bibfield  {journal} {\bibinfo  {journal}
  {arXiv: 1612.08758}\ } (\bibinfo {year} {2016})}\BibitemShut {NoStop}%
\bibitem [{\citenamefont {{Zhang}}\ \emph {et~al.}(2017)\citenamefont
  {{Zhang}}, \citenamefont {{Hess}}, \citenamefont {{Kyprianidis}},
  \citenamefont {{Becker}}, \citenamefont {{Lee}}, \citenamefont {{Smith}},
  \citenamefont {{Pagano}}, \citenamefont {{Potirniche}}, \citenamefont
  {{Potter}}, \citenamefont {{Vishwanath}}, \citenamefont {{Yao}},\ and\
  \citenamefont {{Monroe}}}]{Zhang2016}%
  \BibitemOpen
  \bibfield  {author} {\bibinfo {author} {\bibfnamefont {J.}~\bibnamefont
  {{Zhang}}}, \bibinfo {author} {\bibfnamefont {P.~W.}\ \bibnamefont {{Hess}}},
  \bibinfo {author} {\bibfnamefont {A.}~\bibnamefont {{Kyprianidis}}}, \bibinfo
  {author} {\bibfnamefont {P.}~\bibnamefont {{Becker}}}, \bibinfo {author}
  {\bibfnamefont {A.}~\bibnamefont {{Lee}}}, \bibinfo {author} {\bibfnamefont
  {J.}~\bibnamefont {{Smith}}}, \bibinfo {author} {\bibfnamefont
  {G.}~\bibnamefont {{Pagano}}}, \bibinfo {author} {\bibfnamefont {I.-D.}\
  \bibnamefont {{Potirniche}}}, \bibinfo {author} {\bibfnamefont {A.~C.}\
  \bibnamefont {{Potter}}}, \bibinfo {author} {\bibfnamefont {A.}~\bibnamefont
  {{Vishwanath}}}, \bibinfo {author} {\bibfnamefont {N.~Y.}\ \bibnamefont
  {{Yao}}}, \ and\ \bibinfo {author} {\bibfnamefont {C.}~\bibnamefont
  {{Monroe}}},\ }\href@noop {} {\bibfield  {journal} {\bibinfo  {journal}
  {Nature}\ }\textbf {\bibinfo {volume} {543}},\ \bibinfo {pages} {217}
  (\bibinfo {year} {2017})}\BibitemShut {NoStop}%
\bibitem [{\citenamefont {{Choi}}\ \emph {et~al.}(2016)\citenamefont {{Choi}},
  \citenamefont {{Choi}}, \citenamefont {{Landig}}, \citenamefont {{Kucsko}},
  \citenamefont {{Zhou}}, \citenamefont {{Isoya}}, \citenamefont {{Jelezko}},
  \citenamefont {{Onoda}}, \citenamefont {{Sumiya}}, \citenamefont {{Khemani}},
  \citenamefont {{von Keyserlingk}}, \citenamefont {{Yao}}, \citenamefont
  {{Demler}},\ and\ \citenamefont {{Lukin}}}]{Choi2016}%
  \BibitemOpen
  \bibfield  {author} {\bibinfo {author} {\bibfnamefont {S.}~\bibnamefont
  {{Choi}}}, \bibinfo {author} {\bibfnamefont {J.}~\bibnamefont {{Choi}}},
  \bibinfo {author} {\bibfnamefont {R.}~\bibnamefont {{Landig}}}, \bibinfo
  {author} {\bibfnamefont {G.}~\bibnamefont {{Kucsko}}}, \bibinfo {author}
  {\bibfnamefont {H.}~\bibnamefont {{Zhou}}}, \bibinfo {author} {\bibfnamefont
  {J.}~\bibnamefont {{Isoya}}}, \bibinfo {author} {\bibfnamefont
  {F.}~\bibnamefont {{Jelezko}}}, \bibinfo {author} {\bibfnamefont
  {S.}~\bibnamefont {{Onoda}}}, \bibinfo {author} {\bibfnamefont
  {H.}~\bibnamefont {{Sumiya}}}, \bibinfo {author} {\bibfnamefont
  {V.}~\bibnamefont {{Khemani}}}, \bibinfo {author} {\bibfnamefont
  {C.}~\bibnamefont {{von Keyserlingk}}}, \bibinfo {author} {\bibfnamefont
  {N.~Y.}\ \bibnamefont {{Yao}}}, \bibinfo {author} {\bibfnamefont
  {E.}~\bibnamefont {{Demler}}}, \ and\ \bibinfo {author} {\bibfnamefont
  {M.~D.}\ \bibnamefont {{Lukin}}},\ }\href@noop {} {\bibfield  {journal}
  {\bibinfo  {journal} {Nature}\ }\textbf {\bibinfo {volume} {543}},\ \bibinfo
  {pages} {221} (\bibinfo {year} {2016})}\BibitemShut {NoStop}%
\bibitem [{\citenamefont {{Bairey}}\ \emph {et~al.}(2017)\citenamefont
  {{Bairey}}, \citenamefont {{Refael}},\ and\ \citenamefont
  {{Lindner}}}]{Bairey2017}%
  \BibitemOpen
  \bibfield  {author} {\bibinfo {author} {\bibfnamefont {E.}~\bibnamefont
  {{Bairey}}}, \bibinfo {author} {\bibfnamefont {G.}~\bibnamefont {{Refael}}},
  \ and\ \bibinfo {author} {\bibfnamefont {N.~H.}\ \bibnamefont {{Lindner}}},\
  }\href@noop {} {\bibfield  {journal} {\bibinfo  {journal} {arXiv:
  1702.06208}\ } (\bibinfo {year} {2017})}\BibitemShut {NoStop}%
\bibitem [{\citenamefont {D'Alessio}\ and\ \citenamefont
  {Rigol}(2015)}]{DAlessio2015}%
  \BibitemOpen
  \bibfield  {author} {\bibinfo {author} {\bibfnamefont {L.}~\bibnamefont
  {D'Alessio}}\ and\ \bibinfo {author} {\bibfnamefont {M.}~\bibnamefont
  {Rigol}},\ }\href {http://dx.doi.org/10.1038/ncomms9336} {\bibfield
  {journal} {\bibinfo  {journal} {Nat Commun}\ }\textbf {\bibinfo {volume}
  {6}},\  (\bibinfo {year} {2015})}\BibitemShut {NoStop}%
\bibitem [{\citenamefont {Heinisch}\ and\ \citenamefont
  {Holthaus}(2016)}]{Heinisch2016}%
  \BibitemOpen
  \bibfield  {author} {\bibinfo {author} {\bibfnamefont {C.}~\bibnamefont
  {Heinisch}}\ and\ \bibinfo {author} {\bibfnamefont {M.}~\bibnamefont
  {Holthaus}},\ }\href@noop {} {\bibfield  {journal} {\bibinfo  {journal} {J.
  Mod. Opt.}\ ,\ \bibinfo {pages} {1}} (\bibinfo {year} {2016})}\BibitemShut
  {NoStop}%
\bibitem [{\citenamefont {Weinberg}\ \emph {et~al.}(2016)\citenamefont
  {Weinberg}, \citenamefont {Bukov}, \citenamefont {D\'Alessio}, \citenamefont
  {Polkovnikov}, \citenamefont {Vajna},\ and\ \citenamefont
  {Kolodrubetz}}]{Weinberg2016}%
  \BibitemOpen
  \bibfield  {author} {\bibinfo {author} {\bibfnamefont {P.}~\bibnamefont
  {Weinberg}}, \bibinfo {author} {\bibfnamefont {M.}~\bibnamefont {Bukov}},
  \bibinfo {author} {\bibfnamefont {L.}~\bibnamefont {D\'Alessio}}, \bibinfo
  {author} {\bibfnamefont {A.}~\bibnamefont {Polkovnikov}}, \bibinfo {author}
  {\bibfnamefont {S.}~\bibnamefont {Vajna}}, \ and\ \bibinfo {author}
  {\bibfnamefont {M.}~\bibnamefont {Kolodrubetz}},\ }\href@noop {} {\bibfield
  {journal} {\bibinfo  {journal} {arXiv:2016.02229}\ } (\bibinfo {year}
  {2016})}\BibitemShut {NoStop}%
\bibitem [{\citenamefont {{Ho}}\ and\ \citenamefont {{Abanin}}(2016)}]{Ho2016}%
  \BibitemOpen
  \bibfield  {author} {\bibinfo {author} {\bibfnamefont {W.~W.}\ \bibnamefont
  {{Ho}}}\ and\ \bibinfo {author} {\bibfnamefont {D.~A.}\ \bibnamefont
  {{Abanin}}},\ }\href@noop {} {\bibfield  {journal} {\bibinfo  {journal}
  {arXiv:1611.05024}\ } (\bibinfo {year} {2016})}\BibitemShut {NoStop}%
\bibitem [{\citenamefont {Goto}(2016)}]{Goto2016}%
  \BibitemOpen
  \bibfield  {author} {\bibinfo {author} {\bibfnamefont {H.}~\bibnamefont
  {Goto}},\ }\href@noop {} {\bibfield  {journal} {\bibinfo  {journal} {Sci.
  Rep.}\ }\textbf {\bibinfo {volume} {6}},\ \bibinfo {pages} {21686} (\bibinfo
  {year} {2016})}\BibitemShut {NoStop}%
\bibitem [{\citenamefont {{Puri}}\ and\ \citenamefont
  {{Blais}}(2016)}]{Puri2016}%
  \BibitemOpen
  \bibfield  {author} {\bibinfo {author} {\bibfnamefont {S.}~\bibnamefont
  {{Puri}}}\ and\ \bibinfo {author} {\bibfnamefont {A.}~\bibnamefont
  {{Blais}}},\ }\href@noop {} {\bibfield  {journal} {\bibinfo  {journal}
  {arXiv: 1605.09408}\ } (\bibinfo {year} {2016})}\BibitemShut {NoStop}%
\bibitem [{\citenamefont {Larsen}\ and\ \citenamefont
  {Bloembergen}(1976)}]{Larsen1976}%
  \BibitemOpen
  \bibfield  {author} {\bibinfo {author} {\bibfnamefont {D.~M.}\ \bibnamefont
  {Larsen}}\ and\ \bibinfo {author} {\bibfnamefont {N.}~\bibnamefont
  {Bloembergen}},\ }\href@noop {} {\bibfield  {journal} {\bibinfo  {journal}
  {Opt. Commun.}\ }\textbf {\bibinfo {volume} {17}},\ \bibinfo {pages} {254}
  (\bibinfo {year} {1976})}\BibitemShut {NoStop}%
\bibitem [{\citenamefont {Dykman}\ and\ \citenamefont
  {Fistul}(2005)}]{Dykman2005}%
  \BibitemOpen
  \bibfield  {author} {\bibinfo {author} {\bibfnamefont {M.~I.}\ \bibnamefont
  {Dykman}}\ and\ \bibinfo {author} {\bibfnamefont {M.~V.}\ \bibnamefont
  {Fistul}},\ }\href@noop {} {\bibfield  {journal} {\bibinfo  {journal} {Phys.
  Rev. B}\ }\textbf {\bibinfo {volume} {71}},\ \bibinfo {pages} {140508}
  (\bibinfo {year} {2005})}\BibitemShut {NoStop}%
\bibitem [{\citenamefont {Dykman}(2012)}]{Dykman2012}%
  \BibitemOpen
  \bibfield  {author} {\bibinfo {author} {\bibfnamefont {M.~I.}\ \bibnamefont
  {Dykman}},\ }in\ \href@noop {} {\emph {\bibinfo {booktitle} {Fluctuating
  Nonlinear Oscillators: from Nanomechanics to Quantum Superconducting
  Circuits}}},\ \bibinfo {editor} {edited by\ \bibinfo {editor} {\bibfnamefont
  {M.~I.}\ \bibnamefont {Dykman}}}\ (\bibinfo  {publisher} {OUP, Oxford},\
  \bibinfo {year} {2012})\ pp.\ \bibinfo {pages} {165--197}\BibitemShut
  {NoStop}%
\bibitem [{\citenamefont {Nabors}\ \emph {et~al.}(1990)\citenamefont {Nabors},
  \citenamefont {Yang}, \citenamefont {Day},\ and\ \citenamefont
  {Byer}}]{Nabors1990}%
  \BibitemOpen
  \bibfield  {author} {\bibinfo {author} {\bibfnamefont {C.~D.}\ \bibnamefont
  {Nabors}}, \bibinfo {author} {\bibfnamefont {S.~T.}\ \bibnamefont {Yang}},
  \bibinfo {author} {\bibfnamefont {T.}~\bibnamefont {Day}}, \ and\ \bibinfo
  {author} {\bibfnamefont {R.~L.}\ \bibnamefont {Byer}},\ }\href
  {http://josab.osa.org/abstract.cfm?URI=josab-7-5-815} {\bibfield  {journal}
  {\bibinfo  {journal} {J. Opt. Soc. Am. B}\ }\textbf {\bibinfo {volume} {7}},\
  \bibinfo {pages} {815} (\bibinfo {year} {1990})}\BibitemShut {NoStop}%
\bibitem [{\citenamefont {{Wilson}}\ \emph {et~al.}(2010)\citenamefont
  {{Wilson}}, \citenamefont {{Duty}}, \citenamefont {{Sandberg}}, \citenamefont
  {{Persson}}, \citenamefont {{Shumeiko}},\ and\ \citenamefont
  {{Delsing}}}]{Wilson2010}%
  \BibitemOpen
  \bibfield  {author} {\bibinfo {author} {\bibfnamefont {C.~M.}\ \bibnamefont
  {{Wilson}}}, \bibinfo {author} {\bibfnamefont {T.}~\bibnamefont {{Duty}}},
  \bibinfo {author} {\bibfnamefont {M.}~\bibnamefont {{Sandberg}}}, \bibinfo
  {author} {\bibfnamefont {F.}~\bibnamefont {{Persson}}}, \bibinfo {author}
  {\bibfnamefont {V.}~\bibnamefont {{Shumeiko}}}, \ and\ \bibinfo {author}
  {\bibfnamefont {P.}~\bibnamefont {{Delsing}}},\ }\href@noop {} {\bibfield
  {journal} {\bibinfo  {journal} {Phys. Rev. Lett.}\ }\textbf {\bibinfo
  {volume} {105}},\ \bibinfo {pages} {233907} (\bibinfo {year}
  {2010})}\BibitemShut {NoStop}%
\bibitem [{\citenamefont {Lin}\ \emph {et~al.}(2014)\citenamefont {Lin},
  \citenamefont {Inomata}, \citenamefont {Koshino}, \citenamefont {Oliver},
  \citenamefont {Nakamura}, \citenamefont {Tsai},\ and\ \citenamefont
  {Yamamoto}}]{Lin2014}%
  \BibitemOpen
  \bibfield  {author} {\bibinfo {author} {\bibfnamefont {Z.}~\bibnamefont
  {Lin}}, \bibinfo {author} {\bibfnamefont {K.}~\bibnamefont {Inomata}},
  \bibinfo {author} {\bibfnamefont {K.}~\bibnamefont {Koshino}}, \bibinfo
  {author} {\bibfnamefont {W.}~\bibnamefont {Oliver}}, \bibinfo {author}
  {\bibfnamefont {Y.}~\bibnamefont {Nakamura}}, \bibinfo {author}
  {\bibfnamefont {J.}~\bibnamefont {Tsai}}, \ and\ \bibinfo {author}
  {\bibfnamefont {T.}~\bibnamefont {Yamamoto}},\ }\href@noop {} {\bibfield
  {journal} {\bibinfo  {journal} {Nat Commun}\ }\textbf {\bibinfo {volume}
  {5}},\ \bibinfo {pages} {4480} (\bibinfo {year} {2014})}\BibitemShut
  {NoStop}%
\bibitem [{\citenamefont {Haroche}\ and\ \citenamefont
  {Raimond}(2006)}]{Haroche2006}%
  \BibitemOpen
  \bibfield  {author} {\bibinfo {author} {\bibfnamefont {S.}~\bibnamefont
  {Haroche}}\ and\ \bibinfo {author} {\bibfnamefont {J.~M.}\ \bibnamefont
  {Raimond}},\ }\href@noop {} {\emph {\bibinfo {title} {Exploring the Quantum:
  Atoms, Cavities, and Photons}}}\ (\bibinfo  {publisher} {Oxford Univ.
  Press},\ \bibinfo {address} {Oxford},\ \bibinfo {year} {2006})\BibitemShut
  {NoStop}%
\bibitem [{\citenamefont {Marthaler}\ and\ \citenamefont
  {Dykman}(2007)}]{Marthaler2007a}%
  \BibitemOpen
  \bibfield  {author} {\bibinfo {author} {\bibfnamefont {M.}~\bibnamefont
  {Marthaler}}\ and\ \bibinfo {author} {\bibfnamefont {M.~I.}\ \bibnamefont
  {Dykman}},\ }\href@noop {} {\bibfield  {journal} {\bibinfo  {journal} {Phys.
  Rev. A}\ }\textbf {\bibinfo {volume} {76}},\ \bibinfo {pages} {010102R}
  (\bibinfo {year} {2007})}\BibitemShut {NoStop}%
\bibitem [{\citenamefont {Zhang}\ \emph {et~al.}(2017)\citenamefont {Zhang},
  \citenamefont {Gosner}, \citenamefont {Ankerhold}, \citenamefont {Girvin},\
  and\ \citenamefont {Dykman}}]{Zhang2017}%
  \BibitemOpen
  \bibfield  {author} {\bibinfo {author} {\bibfnamefont {Y.}~\bibnamefont
  {Zhang}}, \bibinfo {author} {\bibfnamefont {J.}~\bibnamefont {Gosner}},
  \bibinfo {author}{\bibfnamefont {S.~M.}\ \bibnamefont {Girvin}}, \bibinfo
  {author} {\bibfnamefont {J.}~\bibnamefont {Ankerhold}}, \ and\ \bibinfo
  {author} {\bibfnamefont {M.}~\bibnamefont {Dykman}},\ }\href@noop {}
  {\bibfield  {journal} {\bibinfo  {journal} {arXiv:1702.07931}\ } (\bibinfo
  {year} {2017})}\BibitemShut {NoStop}%
\bibitem [{\citenamefont {Mandel}\ and\ \citenamefont
  {Wolf}(1995)}]{Mandel1995}%
  \BibitemOpen
  \bibfield  {author} {\bibinfo {author} {\bibfnamefont {L.}~\bibnamefont
  {Mandel}}\ and\ \bibinfo {author} {\bibfnamefont {E.}~\bibnamefont {Wolf}},\
  }\href@noop {} {\emph {\bibinfo {title} {Optical Coherence and Quantum
  Optics}}}\ (\bibinfo  {publisher} {Cambirdge University Press},\ \bibinfo
  {year} {Cambridge, 1995})\BibitemShut {NoStop}%
\bibitem [{\citenamefont {Marthaler}\ and\ \citenamefont
  {Dykman}(2006)}]{Marthaler2006}%
  \BibitemOpen
  \bibfield  {author} {\bibinfo {author} {\bibfnamefont {M.}~\bibnamefont
  {Marthaler}}\ and\ \bibinfo {author} {\bibfnamefont {M.~I.}\ \bibnamefont
  {Dykman}},\ }\href@noop {} {\bibfield  {journal} {\bibinfo  {journal} {Phys.
  Rev. A}\ }\textbf {\bibinfo {volume} {73}},\ \bibinfo {pages} {042108}
  (\bibinfo {year} {2006})}\BibitemShut {NoStop}%
\bibitem [{Note1()}]{Note1}%
  \BibitemOpen
  \bibinfo {note} {The oscillator is characterized also by a much longer rate,
  which is related to the dissipation-induced transitions between the wells in
  Fig.~1.}\BibitemShut {Stop}%
\bibitem [{\citenamefont {Dykman}(1975)}]{Dykman1975}%
  \BibitemOpen
  \bibfield  {author} {\bibinfo {author} {\bibfnamefont {M.~I.}\ \bibnamefont
  {Dykman}},\ }\href@noop {} {\bibfield  {journal} {\bibinfo  {journal} {Zh.
  Eksp. Teor. Fiz.}\ }\textbf {\bibinfo {volume} {68}},\ \bibinfo {pages}
  {2082} (\bibinfo {year} {1975})}\BibitemShut {NoStop}%
\bibitem [{\citenamefont {Whittaker}\ and\ \citenamefont
  {Watson}(1990)}]{Whittaker1990}%
  \BibitemOpen
  \bibfield  {author} {\bibinfo {author} {\bibfnamefont {E.~T.}\ \bibnamefont
  {Whittaker}}\ and\ \bibinfo {author} {\bibfnamefont {G.~N.}\ \bibnamefont
  {Watson}},\ }\href@noop {} {\emph {\bibinfo {title} {A Course in Modern
  Analysis}}},\ \bibinfo {edition} {4th}\ ed.\ (\bibinfo  {publisher}
  {Cambirdge University Press},\ \bibinfo {year} {1990})\BibitemShut {NoStop}%
\bibitem [{\citenamefont {Abramowitz}\ and\ \citenamefont
  {Stegun}(1972)}]{Abramowitz1972}%
  \BibitemOpen
  \bibinfo {editor} {\bibfnamefont {M.}~\bibnamefont {Abramowitz}}\ and\
  \bibinfo {editor} {\bibfnamefont {I.~A.}\ \bibnamefont {Stegun}},\ eds.,\
  \href@noop {} {\emph {\bibinfo {title} {Handbook of Mathematical Functions
  with Formulas, Graphs, and Mathematical Table}}}\ (\bibinfo  {publisher}
  {Dover Publications, Inc.},\ \bibinfo {year} {1972})\BibitemShut {NoStop}%
\end{thebibliography}

%

\end{document}